\documentclass[
 reprint,
superscriptaddress,
 amsmath,amssymb,
 aps,prl
]{revtex4-2}
\usepackage{graphicx}
\usepackage{dcolumn}
\usepackage{bm}
\usepackage{hyperref,xcolor,times,setspace}

\definecolor{linkblue}{rgb}{0, 0, 1}
\hypersetup{
	colorlinks=true,
	linkcolor=linkblue, 
	citecolor=linkblue,   
	urlcolor=linkblue,
}

\begin{document}

\title{Characterizing Biphoton Spatial Wave Function Dynamics with Quantum Wavefront Sensing}

\author{Yi~Zheng}
\affiliation{CAS Key Laboratory of Quantum Information, University of Science and Technology of China, Hefei 230026, China}
\affiliation{CAS Center for Excellence in Quantum Information and Quantum Physics, University of Science and Technology of China, Hefei 230026, China}
	
\author{Zhao-Di~Liu}
\email{zdliu@ustc.edu.cn}
\affiliation{CAS Key Laboratory of Quantum Information, University of Science and Technology of China, Hefei 230026, China}
\affiliation{CAS Center for Excellence in Quantum Information and Quantum Physics, University of Science and Technology of China, Hefei 230026, China}
	
\author{Rui-Heng~Miao}
\affiliation{CAS Key Laboratory of Quantum Information, University of Science and Technology of China, Hefei 230026, China}
\affiliation{CAS Center for Excellence in Quantum Information and Quantum Physics, University of Science and Technology of China, Hefei 230026, China}
\affiliation{Hefei National Laboratory, University of Science and Technology of China, Hefei 230088, China}

\author{Jin-Ming~Cui}
\affiliation{CAS Key Laboratory of Quantum Information, University of Science and Technology of China, Hefei 230026, China}
\affiliation{CAS Center for Excellence in Quantum Information and Quantum Physics, University of Science and Technology of China, Hefei 230026, China}
\affiliation{Hefei National Laboratory, University of Science and Technology of China, Hefei 230088, China}

\author{Mu~Yang}
\affiliation{CAS Key Laboratory of Quantum Information, University of Science and Technology of China, Hefei 230026, China}
\affiliation{CAS Center for Excellence in Quantum Information and Quantum Physics, University of Science and Technology of China, Hefei 230026, China}

\author{Xiao-Ye~Xu}
\affiliation{CAS Key Laboratory of Quantum Information, University of Science and Technology of China, Hefei 230026, China}
\affiliation{CAS Center for Excellence in Quantum Information and Quantum Physics, University of Science and Technology of China, Hefei 230026, China}
\affiliation{Hefei National Laboratory, University of Science and Technology of China, Hefei 230088, China}

\author{Jin-Shi~Xu}
\email{jsxu@ustc.edu.cn}
\affiliation{CAS Key Laboratory of Quantum Information, University of Science and Technology of China, Hefei 230026, China}
\affiliation{CAS Center for Excellence in Quantum Information and Quantum Physics, University of Science and Technology of China, Hefei 230026, China}
\affiliation{Hefei National Laboratory, University of Science and Technology of China, Hefei 230088, China}

\author{Chuan-Feng~Li}
\email{cfli@ustc.edu.cn}
\affiliation{CAS Key Laboratory of Quantum Information, University of Science and Technology of China, Hefei 230026, China}
\affiliation{CAS Center for Excellence in Quantum Information and Quantum Physics, University of Science and Technology of China, Hefei 230026, China}
\affiliation{Hefei National Laboratory, University of Science and Technology of China, Hefei 230088, China}
	
\author{Guang-Can~Guo}
\affiliation{CAS Key Laboratory of Quantum Information, University of Science and Technology of China, Hefei 230026, China}
\affiliation{CAS Center for Excellence in Quantum Information and Quantum Physics, University of Science and Technology of China, Hefei 230026, China}
\affiliation{Hefei National Laboratory, University of Science and Technology of China, Hefei 230088, China}

\date{\today}

\begin{abstract}
With an extremely high dimensionality, the spatial degree of freedom of entangled photons is a key tool for quantum foundation and applied quantum techniques. To fully utilize the feature, the essential task is to experimentally characterize the multiphoton spatial wave function including the entangled amplitude and phase information at different evolutionary stages. However, there is no effective method to measure it. Quantum state tomography is costly, and quantum holography requires additional references. Here, we introduce quantum Shack--Hartmann wavefront sensing to perform efficient and reference-free measurement of the biphoton spatial wave function. The joint probability distribution of photon pairs at the back focal plane of a microlens array is measured and used for amplitude extraction and phase reconstruction. In the experiment, we observe that the biphoton amplitude correlation becomes weak while phase correlation shows up during free-space propagation. Our work is a crucial step in quantum physical and adaptive optics and paves the way for characterizing quantum optical fields with high-order correlations or topological patterns.
\end{abstract}

\maketitle

\emph{Introduction.}---The photon is a promising system in fundamental quantum physics and applied quantum techniques \cite{photonrev}. Its spatial degree of freedom is the core of high-dimensional quantum communication and quantum imaging \cite{contrev}, including ghost imaging \cite{ghostimaging}, imaging with undetected photons \cite{Lemos2014}, image distillation \cite{distill}, and superresolution imaging \cite{PhysRevA.79.013827,He2023}. Developing experimental methods to characterize the entangled multiphoton states is a basic task \cite{PhysRevLett.123.150402}. For the spatial state (sometimes called the wave function \cite{Lundeen2011}), direct coincidence counting can reveal only the joint probability distribution (JPD), i.e., the squared modulus of the wave function \cite{PhysRevLett.92.210403}, while effective phase measurement methods will become useful in various applications like information encoding \cite{Wang2012}, biomedical phase imaging \cite{Park2018}, and aberration cancellation. Moreover, photons are prone to various evolution and modulation processes, so observing the state dynamics is even more exciting and challenging.

Researches on biphoton phase measurement can be inspired by classical methods. A prominent type is holography \cite{GABOR1948}, which needs a reference beam for interference. It has the highest accuracy and is very suitable for characterizing transparent objects. In quantum optics, polarization entanglement enables phase-shifting holography \cite{Yamaguchi:97,Defienne2021,sciadv.abj2155}, and biphoton interference has been employed to measure the biphoton spatial wave function \cite{Zia2023}. However, the reference beam may contain aberrations and is not always available. Starting from Zernike's phase contrast microscopy \cite{Zernikephase}, one type of reference-free method is selecting a part of the unknown field as the reference, including some weak measurement methods \cite{RevModPhys.86.307,Lundeen2011,Shi:15}. Our group used the setup devised by Kocsis \emph{et al.}\ \cite{Kocsis1170} to obtain the phase gradient distribution \cite{Yang2020} for phase reconstruction \cite{Hudgin:77,Southwell:80}, and named it the weak measurement wavefront sensor \cite{Zheng:21,Zheng:22}. Like some shearing methods, it is an interference of the original beam with a slightly displaced one. Then, our group extended it to the multiphoton case \cite{Zheng2023}, which requires JPD measurement of photons. However, weak measurement methods require a high signal-to-noise ratio, which is difficult for biphoton fields in experiments. There are other methods which do not need any reference. State tomography \cite{PhysRevLett.125.090503} requiring a huge number of projection bases is possible but impractical. Here, we consider the celebrated Shack--Hartmann wavefront sensing (SHWS) \cite{Shack01}, which uses a microlens array to project the phase gradient at each aperture to the displacement of the focused spot \cite{Ares:00,Zheng:21}. The Supplemental Material (SM) \cite{sm} introduces basic concepts of SHWS.

In this proof-of-principle work, we introduce and implement quantum SHWS (QSHWS) by measuring the JPD of photon pairs from spontaneous parametric down-conversion (SPDC) \cite{WALBORN201087,Schneeloch_2016} at the focal plane of the microlens array. To our knowledge, this is the first single-measurement and reference-free biphoton phase measurement method. As for the JPD measurement, traditional scanning and coincidence counting method \cite{PhysRevLett.92.210403} is time consuming. In 2018, Defienne \emph{et al.}\ demonstrated that it can be calculated from multiple frames taken by an electron-multiplying charge-coupled device (EMCCD) or single-photon avalanche photodiode (SPAD) array camera \cite{PhysRevLett.120.203604,PhysRevA.98.013841,Ndagano2022}, which we refer to as the multiple frame method. Then, we focus on the biphoton propagation dynamics as an example. Chan \emph{et al.}\ \cite{PhysRevA.75.050101} derived that SPDC photon pairs would be less correlated in position after free-space propagation \cite{Bhattacharjee2022} while phase correlation shows up. We use QSHWS to measure the biphoton state after free-space propagation, and show they agree with theoretical predictions using double-Gaussian approximation \cite{PhysRevLett.92.127903,PhysRevA.75.050101,PhysRevA.95.063836,Schneeloch_2016}. Furthermore, we use a spatial light modulator (SLM) to encode a hyperbolic paraboloid (saddle) phase pattern for detection. Finally, we describe potential applications of QSHWS in biphoton physical \cite{PhysRevLett.94.223601} and adaptive optics \cite{PhysRevLett.121.233601,science.adk7825}.

\emph{Theory.}---We consider photon pairs with a definite polarization and wavelength. Denoting the spatial wave function as $\psi(\boldsymbol{\rho}_1,\boldsymbol{\rho}_2)$ and its phase $\phi(\boldsymbol{\rho}_1,\boldsymbol{\rho}_2)=\operatorname{arg}\psi(\boldsymbol{\rho}_1,\boldsymbol{\rho}_2)$, when the postselected position of photon 1 $\boldsymbol{\rho}_1$ takes all the values inside a microlens aperture $S_1$, the conditional state of photon 2 is generally mixed, described by the reduced density matrix $\hat{\varrho}_2=\int_{S_1}d\boldsymbol{\rho}_1\langle\boldsymbol{\rho}_1|\psi\rangle\langle\psi|\boldsymbol{\rho}_1\rangle$ and its function in the position basis $\varrho_2(\boldsymbol{\rho}'_2,\boldsymbol{\rho}_2)=\langle\boldsymbol{\rho}'_2|\hat{\varrho}_2|\boldsymbol{\rho}_2\rangle=\int_{S_1}d\boldsymbol{\rho}_1\psi(\boldsymbol{\rho}_1,\boldsymbol{\rho}'_2)\psi^\ast(\boldsymbol{\rho}_1,\boldsymbol{\rho}_2)$ (corresponding to the mutual coherence function \cite{PhysRevLett.105.010401,Stoklasa2014,PhysRevLett.127.040402,Zheng:21}). If a single-photon pure state is used, from the spot centroid displacement $\Delta\boldsymbol{\rho}$ from the center of another aperture $S_2$ at the camera, the phase gradient at $S_2$ is calculated by $(2\pi/\lambda)\sin[\arctan(\Delta\boldsymbol{\rho}/f_\mathrm{SH})]$, where $f_\mathrm{SH}$ is the microlens focal length. However, using the mixed state $\hat{\varrho}_2$, the measured phase gradient becomes \cite{Zheng:21,gradientnote} (See SM \cite{sm} for a full derivation)
\begin{align}\label{avgphasgrad}
&\frac{\int_{S_2}d\boldsymbol{\rho}_2\operatorname{Im}\nabla_1\varrho_2(\boldsymbol{\rho}_2,\boldsymbol{\rho}_2)}{\int_{S_2}d\boldsymbol{\rho}_2\varrho_2(\boldsymbol{\rho}_2,\boldsymbol{\rho}_2)}\nonumber\\
=&\frac{\int_{S_1}d\boldsymbol{\rho}_1\int_{S_2}d\boldsymbol{\rho}_2|\psi(\boldsymbol{\rho}_1,\boldsymbol{\rho}_2)|^2\nabla_2\phi(\boldsymbol{\rho}_1,\boldsymbol{\rho}_2)}{\int_{S_1}d\boldsymbol{\rho}_1\int_{S_2}d\boldsymbol{\rho}_2|\psi(\boldsymbol{\rho}_1,\boldsymbol{\rho}_2)|^2},
\end{align}
When the aperture is sufficiently small, $S_1$ and $S_2$ can be approximated by points $\boldsymbol{\rho}_1$ and $\boldsymbol{\rho}_2$, and Eq.\ \eqref{avgphasgrad} becomes the partial phase gradient $\boldsymbol{k}_2(\boldsymbol{\rho}_1,\boldsymbol{\rho}_2)=\nabla_2\phi(\boldsymbol{\rho}_1,\boldsymbol{\rho}_2)$. Exchanging $\boldsymbol{\rho}_1$ and $\boldsymbol{\rho}_2$ yields another partial gradient $\boldsymbol{k}_1(\boldsymbol{\rho}_1,\boldsymbol{\rho}_2)=\nabla_1\phi(\boldsymbol{\rho}_1,\boldsymbol{\rho}_2)$. The joint phase can be reconstructed by line integral \cite{Zheng2023}
\begin{equation}\label{lineint}
\phi(\boldsymbol{\rho}_1,\boldsymbol{\rho}_2)=\int_{(\boldsymbol{0},\boldsymbol{0})}^{(\boldsymbol{\rho}_1,\boldsymbol{\rho}_2)}\boldsymbol{k}_1(\boldsymbol{\rho}'_1,\boldsymbol{\rho}'_2)\cdot d\boldsymbol{\rho}'_1+\boldsymbol{k}_2(\boldsymbol{\rho}'_1,\boldsymbol{\rho}'_2)\cdot d\boldsymbol{\rho}'_2.
\end{equation}
For the amplitude part, with the measured JPD $\Gamma_\mathrm{SH}(\boldsymbol{\rho}_1,\boldsymbol{\rho}_2)$ at the microlens focal plane, assuming each microlens is lossless and its incoming beam does not escape its aperture, the JPD of two apertures $\Gamma(\boldsymbol{\rho}_1,\boldsymbol{\rho}_2)\approx\int_{S_1}d\boldsymbol{\rho}'_1\int_{S_2}d\boldsymbol{\rho}'_2\Gamma_\mathrm{SH}(\boldsymbol{\rho}'_1,\boldsymbol{\rho}'_2)$ (here, $S_1,S_2$ are centered by $\boldsymbol{\rho}_1,\boldsymbol{\rho}_2$ respectively), which equals $\int_{S_1}d\boldsymbol{\rho}'_1\int_{S_2}d\boldsymbol{\rho}'_2|\psi(\boldsymbol{\rho}'_1,\boldsymbol{\rho}'_2)|^2$. The reconstructed wave function is, thus, $\sqrt{\Gamma(\boldsymbol{\rho}_1,\boldsymbol{\rho}_2)}\exp[i\phi(\boldsymbol{\rho}_1,\boldsymbol{\rho}_2)]$.

\begin{figure}[t]
\includegraphics[width=0.48\textwidth]{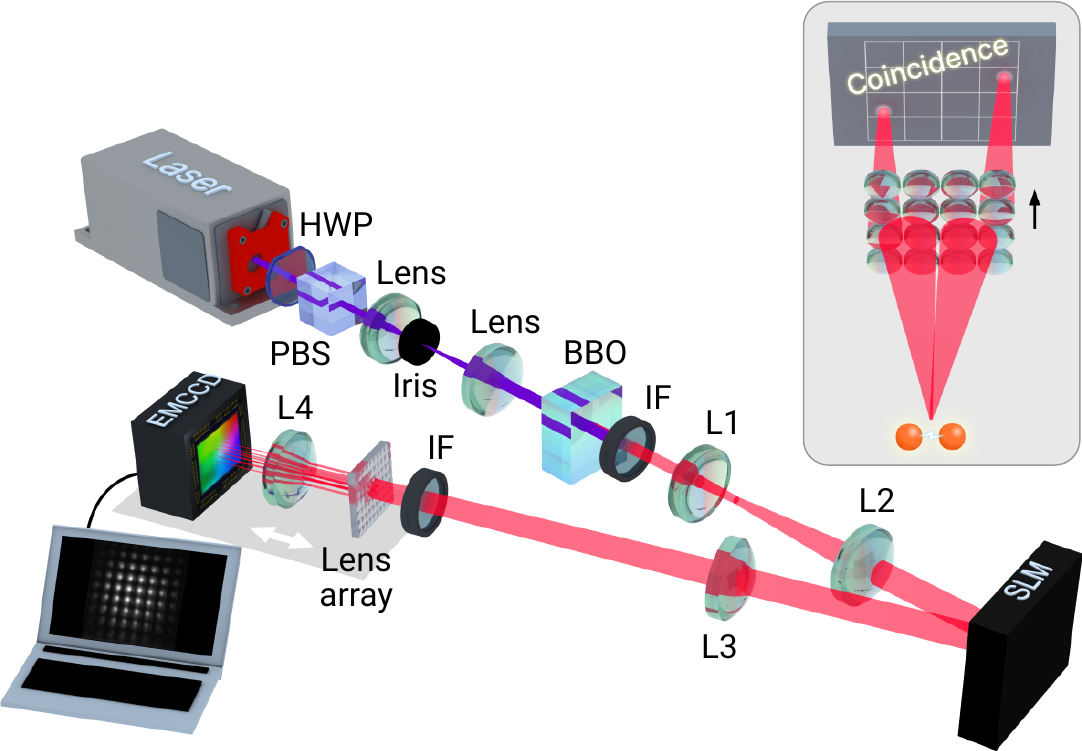}
\caption{\label{setupfig}The experimental setup. A 405-nm, horizontally polarized laser beam is shaped by two lenses and an iris. Then, it pumps a BBO crystal to produce collinear type-I SPDC photon pairs, and is filtered out by a long-pass interference filter (IF). Two Fourier lenses (L1 and L2) project the photons from the BBO to the spatial light modulator (SLM). Then another Fourier lens (L3) projects the photons to the microlens array. A bandpass IF at $(810\pm5)\,\mathrm{nm}$ selects degenerate SPDC photon pairs. An imaging lens (L4) images the photons from the focal plane of the microlens array to the EMCCD sensor. The microlens array, L4, and EMCCD can be displaced by a distance. The inset shows the basic idea of quantum Shack--Hartmann wavefront sensing.}
\end{figure}

If the two photons are indistinguishable $\psi(\boldsymbol{\rho}_1,\boldsymbol{\rho}_2)=\psi(\boldsymbol{\rho}_2,\boldsymbol{\rho}_1)$, only one phase gradient distribution can be measured. Without loss of generality, let it be $\boldsymbol{k}_1(\boldsymbol{\rho}_1,\boldsymbol{\rho}_2)$. Then, $\boldsymbol{k}_2(\boldsymbol{\rho}_1,\boldsymbol{\rho}_2)=\boldsymbol{k}_1(\boldsymbol{\rho}_2,\boldsymbol{\rho}_1)$, and the reconstructed $\phi(\boldsymbol{\rho}_1,\boldsymbol{\rho}_2)$ has exchange symmetry.

\emph{Experimental setup.}---The experimental setup is shown in Fig.\ \ref{setupfig}. A pump laser at $405~\mathrm{nm}$ is incident on a $\beta$-barium borate (BBO) crystal. Degenerate near-collinear type-I SPDC photon pairs pass through two Fourier lenses, get reflected by a SLM, pass through another Fourier lens, and arrive at the microlens array. The width of each microlens is $0.3~\mathrm{mm}$. An imaging lens is inserted at the midpoint of the microlens array and the EMCCD sensor. See SM \cite{sm} for details.

\emph{Data processing.}---In one measurement, the EMCCD takes multiple frames, and a threshold determines whether pixels of each frame have at least one photon. In the multiple frame method, the JPD is calculated by the covariance of the counts at two pixels, which, however, suits only for biphoton states with narrow conditional probability distributions (CPDs, with the position of one photon given) due to its signal-to-noise ratio \cite{PhysRevA.98.013841,Bhattacharjee2022}. Also, photon intensity or sensor efficiency fluctuation results in a positive background in the calculated JPD, so a successive frame formula is mainly used in subsequent works \cite{distill,Defienne2021,science.adk7825}, which will cause more noise in the JPD \cite{He2023}, and the background removal is not effective in our experiments. We develop a brightness separation formula where frames registering too many or few photons are discarded, and the background is basically eliminated.

\begin{figure}[t]
\includegraphics[width=0.48\textwidth]{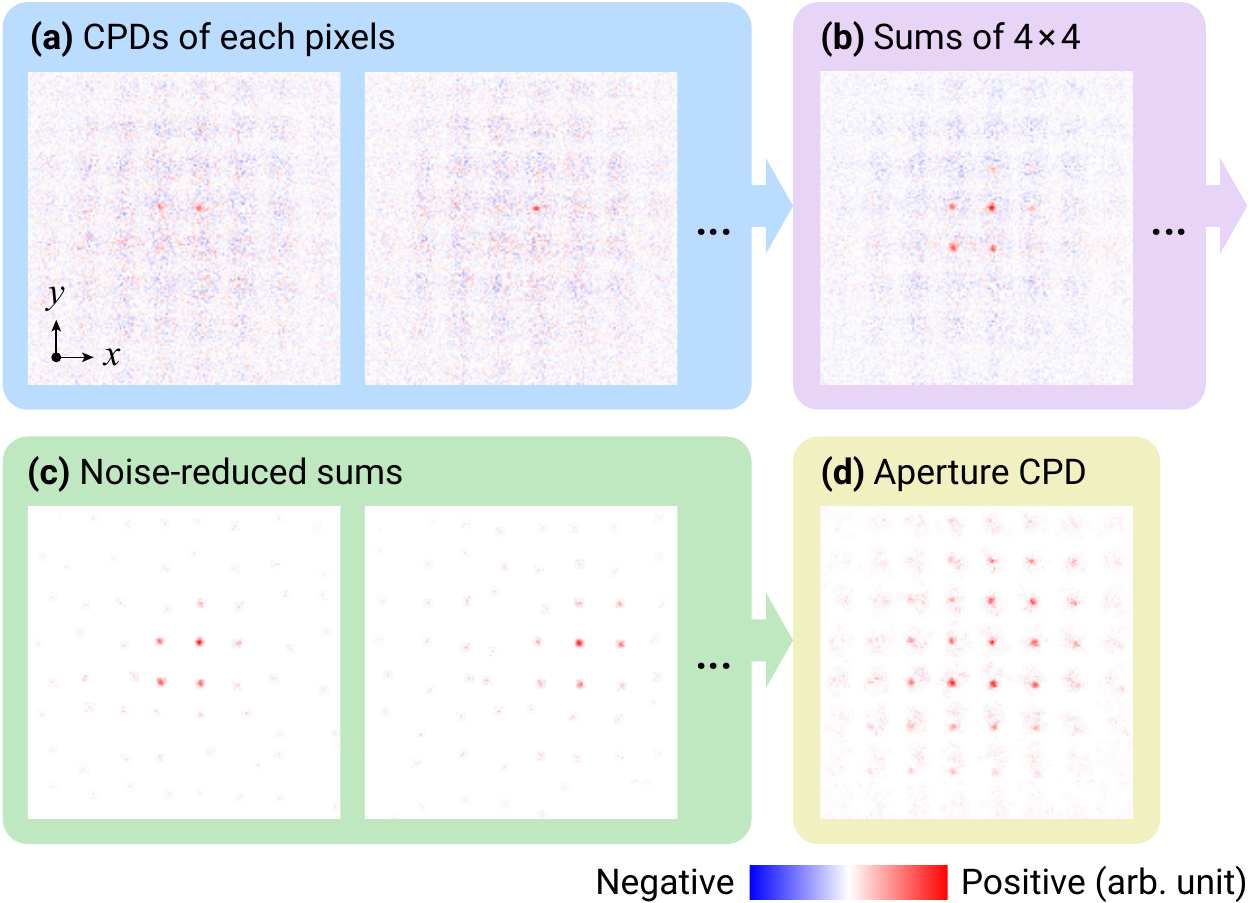}
\caption{\label{prinfig}Calculation of the aperture CPD. (a) The single-pixel CPDs from the multiple frame method. (b) The sum of $16$ CPDs of a $4\times4$-pixel segment in (a). (c) Noise-reduced distributions from (b). (d) The sum of all the distributions in (c), which is the aperture CPD. The $7$-cm propagation data are used as an example. See SM \cite{sm} for more distribution graphs.}
\end{figure}

Single-pixel CPDs of pixels within an aperture are summed to obtain the aperture CPD, but direct summation will distort the result. We choose to divide the aperture into several $4\times4$-pixel segments, sum CPDs over each segment first, and reduce the noise before being summed for the final aperture CPD. Fig.\ \ref{prinfig} illustrates the processes. Finally, the centroid positions of the spots are calculated from the aperture CPD and converted to the phase gradient distribution.

An algorithm is required to realize the phase reconstruction in Eq.\ \eqref{lineint}. For the four-dimensional phase distribution, the traditional zonal and modal methods \cite{Southwell:80} which require large matrix operations is difficult to implement. We adapt the method based on random point spreading and averaging in Ref.\ \cite{Zheng2023}. SM \cite{sm} provides all details of data processing.

\begin{figure}[t]
\includegraphics[width=.48\textwidth]{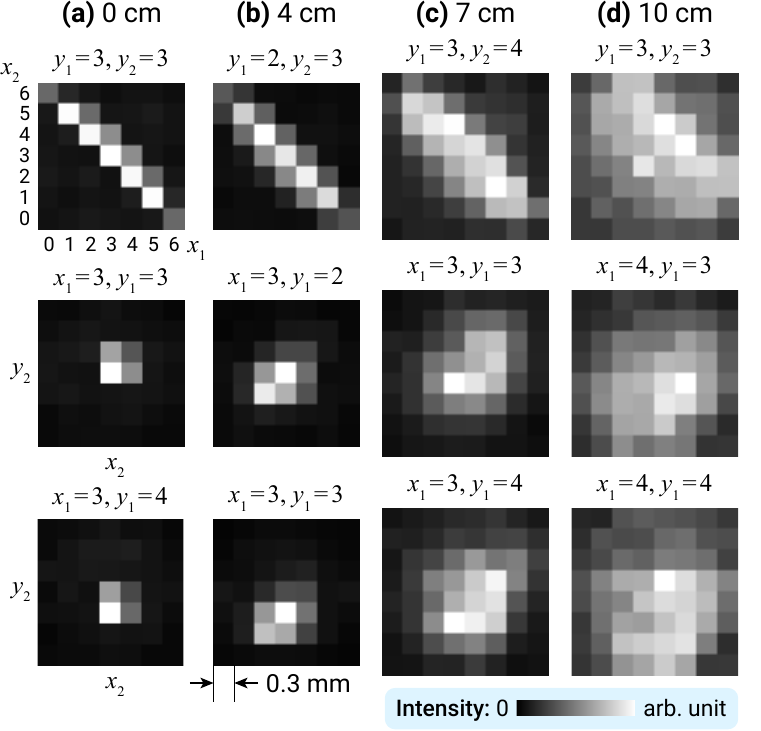}
\caption{\label{anticorfig}Amplitude correlation and CPDs of SPDC photon pairs with (a) no propagation; (b) $4$-cm propagation; (c) $7$-cm; and (d) $10$-cm. The first line is the $x_1,x_2$ distribution when the indices (starting from $0$) of $y_1,y_2$ are given. The other two are CPDs of given apertures $x_1,y_1$.}
\end{figure}

\emph{Biphoton propagation dynamics.}---Using the microlens array, the biphoton position anticorrelation after three Fourier lenses with the SLM off can be observed easily, as shown in Fig.\ \ref{anticorfig}(a). The anticorrelation is not aperture to aperture, because its center is not the center, a side midpoint, or a vertex of a microlens aperture.

However, the phase cannot be reconstructed without magnification for biphoton states with strong position correlation (see Discussion). To demonstrate the phase measurement, we need states with wider CPDs. Because of the immaturity of arbitrary two-photon interaction and modulation, more general correlated phase patterns cannot be generated experimentally. We choose to measure the correlated phase of propagated SPDC photon pairs. Before propagation, their wave function can be approximated by a double-Gaussian function (unnormalized) \cite{PhysRevLett.92.127903,PhysRevA.75.050101,PhysRevA.95.063836,Schneeloch_2016}:
\begin{equation}\label{dGwf}
    \psi_\mathrm{dG}(\boldsymbol{\rho}_1,\boldsymbol{\rho}_2)=\exp\left(-\frac{|\boldsymbol{\rho}_1+\boldsymbol{\rho}_2|^2}{4\sigma_+^2}-\frac{|\boldsymbol{\rho}_1-\boldsymbol{\rho}_2|^2}{4\sigma_-^2}\right).
\end{equation}
In the momentum space (angular spectrum \cite{FourierOptics}), it is the Fourier transform of Eq.\ \eqref{dGwf}:
\begin{equation}\label{dGwfk}
    \tilde{\psi}_\mathrm{dG}(\boldsymbol{q}_1,\boldsymbol{q}_2)=\exp\left(-\frac{\sigma_+^2|\boldsymbol{q}_1+\boldsymbol{q}_2|^2}{4}-\frac{\sigma_-^2|\boldsymbol{q}_1-\boldsymbol{q}_2|^2}{4}\right).
\end{equation}
After propagating a distance $z$, a phase $\exp[-iz(|\boldsymbol{q}_1|^2+|\boldsymbol{q}_2|^2)/(2k)]$ is added to the angular spectrum and, thus, $\sigma_\pm^2\to\sigma_\pm^2+iz/k$, where $k=2\pi/\lambda$ and $\lambda=810~\mathrm{nm}$. Substituting the new complex $\sigma_\pm^2$ into Eq.\ \eqref{dGwf} yields the new wave function with phase correlation. When $z=k\sigma_+\sigma_-$, the photons have no amplitude correlation, and entanglement exists only in the phase \cite{PhysRevA.75.050101}. The amplitude correlation direction switches when $z>k\sigma_+\sigma_-$.

Previously, the biphoton JPD at intermediate distances cannot be measured using the multiple frame method \cite{Bhattacharjee2022}. Now, we reconstruct the wave functions of SPDC photon pairs propagating $4$, $7$, and $10~\mathrm{cm}$. Data of farther distances are still limited by the signal-to-noise ratio. The position correlation patterns and selected CPDs are shown in Figs.\ \ref{anticorfig}(b)--\ref{anticorfig}(d), where the anticorrelation becomes weaker and CPDs become wider as $z$ increases.

\begin{figure}[t]
\includegraphics[width=.48\textwidth]{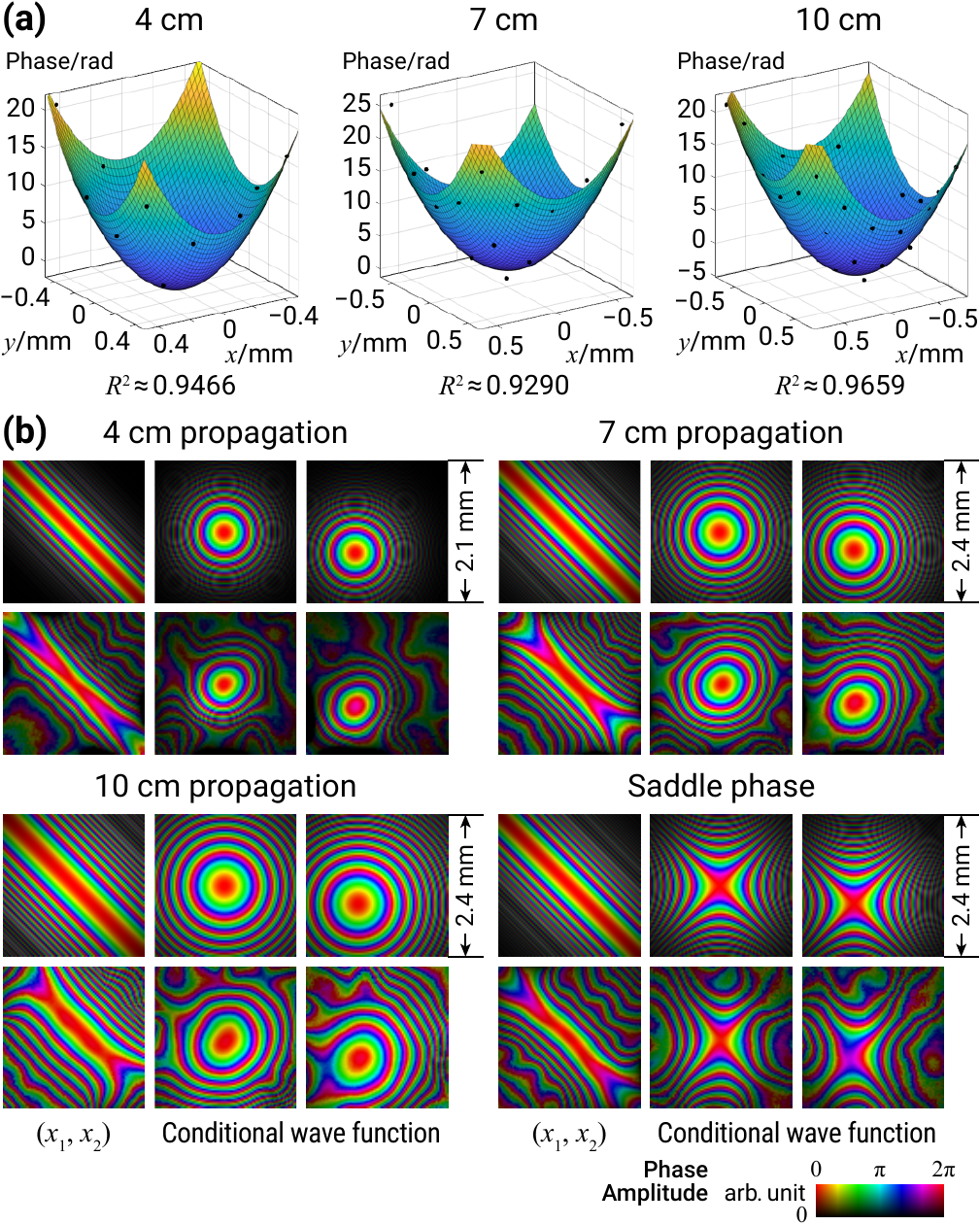}
\caption{\label{difffig}(a) The reconstructed phase distributions of conditional wave functions without interpolation. The given points for the three conditional wave functions are indices $(3,3)$, $(4,4)$, and $(4,3)$ respectively. Dots are the reconstruction results, and curves are theoretical phases with fitted displacement values. Dots away from zero are omitted, as their CPDs are smaller and vulnerable to errors. The $R^2$ values are given. (b) Theoretical double-Gaussian wave functions and experimentally reconstructed ones from interpolated phase gradients. For each biphoton state, the conditional wave function of $x_1,x_2$ with $y_1,y_2$ fixed at the center and two conditional wave functions of $x_2,y_2$ with two given $(x_1,y_1)$ points (the first point is the center, and the second one is on the top right side, $0.3\sqrt{2}~\mathrm{mm}$ from the center) are shown. See SM \cite{sm} for more conditional wave functions.}
\end{figure}

The theoretical phases of conditional wave functions are approximately spherical waves at twice the actual propagation distances, which can be explained in the next section. First, we do not interpolate the phase gradient distribution, and the reconstructed phase of conditional wave functions at a near-center aperture are fitted by phases of spherical waves at distances $8$, $14$, and $20~\mathrm{cm}$ with displacements to be determined as shown in Fig.\ \ref{difffig}(a). The coefficients of determination ($R^2$) characterizing the similarity are also given, showing the reconstructed phases fit well with theory where the JPD values are higher (otherwise, the noise in the aperture CPDs distorts the phase gradient). As the phases of the wave functions are quadratic, their gradients are linear functions of the coordinates, valid for linear interpolation.

Then, we reconstruct wave functions from the interpolated ($10$ times) JPDs of two apertures and gradient distributions. Selected slices of them are shown in Fig.\ \ref{difffig}(b), where the theoretical counterparts from Eq.\ \eqref{dGwf} are also plotted ($\sigma_+\approx13.43~\textrm{\textmu m}$, $\sigma_-\approx1.143~\mathrm{mm}$. See SM \cite{sm} for the calculation, which includes Ref.\ \cite{10.1063/1.339536} about the refractive indices of BBO). Without interpolation, the phase patterns may be difficult to recognize from phase values modulo $2\pi$. In the conditional wave functions, the center of the paraboloid phase pattern is anticorrelated with the given point. In regions where JPD values are higher, the reconstructed wave functions fit well with the theory. If these wave functions are Fourier transformed, the positive momentum correlation and the global paraboloid phase can be observed, as shown in SM \cite{sm}.

Note that a classically correlated photon pair has almost no momentum correlation, so measuring the JPD in the momentum space can tell whether they are entangled \cite{PhysRevLett.92.210403,Courme:23}. In SM \cite{sm}, we give a form of mixed two-photon state with position anticorrelation and briefly analyze its propagation and the measured values, compared with the entangled state above.

\emph{Phase modulation.}---If a phase pattern $\Phi(\boldsymbol{\rho})$ is added on the SLM, the biphoton wave function at the SLM is multiplied by $\exp[i\Phi(\boldsymbol{\rho}_1)+i\Phi(\boldsymbol{\rho}_2)]$. Letting the Fourier transform of $\exp[i\Phi(\boldsymbol{\rho})]$ be $G(\boldsymbol{\rho})=\int d\boldsymbol{\rho}'\exp[i\Phi(\boldsymbol{\rho}')-ik\boldsymbol{\rho}\cdot\boldsymbol{\rho}'/f_3]$, where $f_3$ is the focal length of the third Fourier lens, the wave function at the microlens array becomes $\psi(\boldsymbol{\rho}_1,\boldsymbol{\rho}_2)\ast[G(\boldsymbol{\rho}_1)G(\boldsymbol{\rho}_2)]$. Assuming the photon pairs are originally perfectly anticorrelated $\psi(\boldsymbol{\rho}_1,\boldsymbol{\rho}_2)=\exp(-|\boldsymbol{\rho}_1|^2/\sigma_-^2)\delta(\boldsymbol{\rho}_1+\boldsymbol{\rho}_2)$, it is
\begin{equation}\label{conveqn}
    \left[G(\boldsymbol{\rho}_2)\exp\left(-\frac{|\boldsymbol{\rho}_1-\boldsymbol{\rho}_2|^2}{\sigma_-^2}\right)\right]\ast^{(\boldsymbol{\rho}_2)}G(\boldsymbol{\rho}_1+\boldsymbol{\rho}_2),
\end{equation}
where the convolution performs only on $\boldsymbol{\rho}_2$ (See SM \cite{sm} for its derivation). If $\sigma_-\to+\infty$ (the anticorrelated Einstein--Podolsky--Rosen state \cite{EPRpaper}), the conditional wave function with a given $\boldsymbol{\rho}_1$ is $G(\boldsymbol{\rho}_2)\ast^{(\boldsymbol{\rho}_2)}G(\boldsymbol{\rho}_1+\boldsymbol{\rho}_2)$, which is the Fourier transform of $\exp[i2\Phi(\boldsymbol{\rho}_2)]$ displaced by $-\boldsymbol{\rho}_1$.

Adding a paraboloid phase $\Phi(\boldsymbol{\rho})=-kz|\boldsymbol{\rho}|^2/(2f_3^2)$ on the SLM can simulate the propagation of $z$. The Fourier transform of $\exp[i2\Phi(\boldsymbol{\rho})]$ is a spherical wave at $2z$ in classical optics, and, thus, the conditional wave functions take this form. If a saddle phase $-a(x^2-y^2)$ is added, it propagates forward in the $x$ direction and backward in $y$, resulting in the saddle phase in the conditional wave functions. We choose the forward and backward distance to be $7$ cm. In the reconstructed phase as shown in Fig.\ \ref{difffig}(b), the magnitudes at the $x$ and $y$ directions are slightly different, which is possibly caused by a slight wavefront curvature of the pump beam, the obliquity of the SLM, or the deviation of the imaging lens magnification factor from $-1$. As the latter two can be eliminated, this method may be used to detect the low-order phase added to photon pairs correlated in position.

\emph{Discussion.}---In this work, we introduced the quantum Shack--Hartmann wavefront sensing method to reconstruct the biphoton spatial wave function, developed the data processing algorithms, and performed the experiments of biphoton propagation dynamics and saddle phase modulation. This method requires only one measurement and is reference-free, but the spatial resolution is the major shortcoming. Amplitude and phase patterns with higher spatial complexities cannot be detected without magnification. Although the CPD width limit in the multiple frame method is greatly lowered with the microlens array, the signal-to-noise ratio still hinders the detection of far wider CPDs, which may be overcome by a time-stamping camera like Tpx3Cam \cite{Courme:23,Zia2023}. Recently, another wavefront sensor based on spatial masking and diffraction with a high resolution was invented \cite{yi2021angle}, which may be considered in quantum wavefront sensing in the future.

Classical SHWS has been widely used in fields like astronomy, laser optics, and biomedical imaging, while QSHWS can be applied in various tasks in the emerging field of adaptive quantum optics \cite{PhysRevLett.121.233601,science.adk7825}. The phase information that lies in the second-order correlation can be assisted in astronomical observation, multiphoton physical optics \cite{PhysRevLett.94.223601}, and free-space quantum communication against atmospheric turbulence. When studying multiphoton interaction or modulation media like nonlinear crystals and atom ensembles \cite{doi:10.1126/sciadv.adf9161}, the correlated phase aberration cannot be measured classically, while QSHWS may be the solution.

Theoretically, the single-pixel CPD shown in Fig.\ \ref{prinfig}(a) provides more information of the biphoton field, linking the combination of the position and phase gradient of one photon to that of the other. For example, it reveals some properties of mixed states \cite{PhysRevLett.105.010401,Stoklasa2014}, and might be used for information encoding and secret sharing. Although the photon pairs are indistinguishable in our experiments, they can be spatially separated or have different wavelengths. This method is not limited to two photons \cite{Zheng2023}. If the JPD of more photons can be measured, the multiphoton wave function can also be reconstructed, leading to the measurement of higher-order correlations. Finally, this wavefront sensing technique can also be used in other continuous-variable degrees of freedom, such as frequency, by converting them into spatial entanglement \cite{Liu2018,PhysRevA.102.062208}, as well as combining topology and entanglement \cite{PhysRevX.12.031022} in quantum optics.

\begin{acknowledgments}
\vspace{0.5em}
This work was supported by the Innovation Program for Quantum Science and Technology (No.\ 2021ZD0301200 and No.\ 2021ZD0301400), National Natural Science Foundation of China (No.\ 62005263, No.\ 11821404, No.\ U19A2075, and No.\ 92365205), Anhui Initiative in Quantum Information Technologies (No.\ AHY060300), the Fundamental Research Funds for the Central Universities (No.\ WK2030000069), and the China Postdoctoral Science Foundation (No.\ 478 2020M671862).
\vspace{0.5em}

Y.Z. and Z.-D.L. contributed equally to this letter.
\end{acknowledgments}

\clearpage
\newpage
\begin{spacing}{1.02}
\setcounter{page}{1}
\appendix
\setcounter{equation}{0}
\setcounter{figure}{0}
\renewcommand{\thefigure}{S\arabic{figure}}
\renewcommand{\theequation}{S\arabic{equation}}
\onecolumngrid
\renewcommand{\appendixname}{Section}

\begin{center}
    \hypertarget{smpage}{\textbf{\large{Characterizing Biphoton Spatial Wave Function Dynamics with Quantum Wavefront Sensing: Supplemental Material}}}
    
    \vspace{5mm}
    {\small Yi~Zheng, Zhao-Di~Liu, Rui-Heng~Miao, Jin-Ming~Cui, Mu~Yang, Xiao-Ye Xu, Jin-Shi~Xu, Chuan-Feng~Li and Guang-Can~Guo}
\end{center}

\vspace{1em}
\centerline{\bf A.\;Concepts of classical Shack--Hartmann wavefront sensing}
\vspace{1em}

\begin{figure*}[h]
\includegraphics[width=0.26\textwidth]{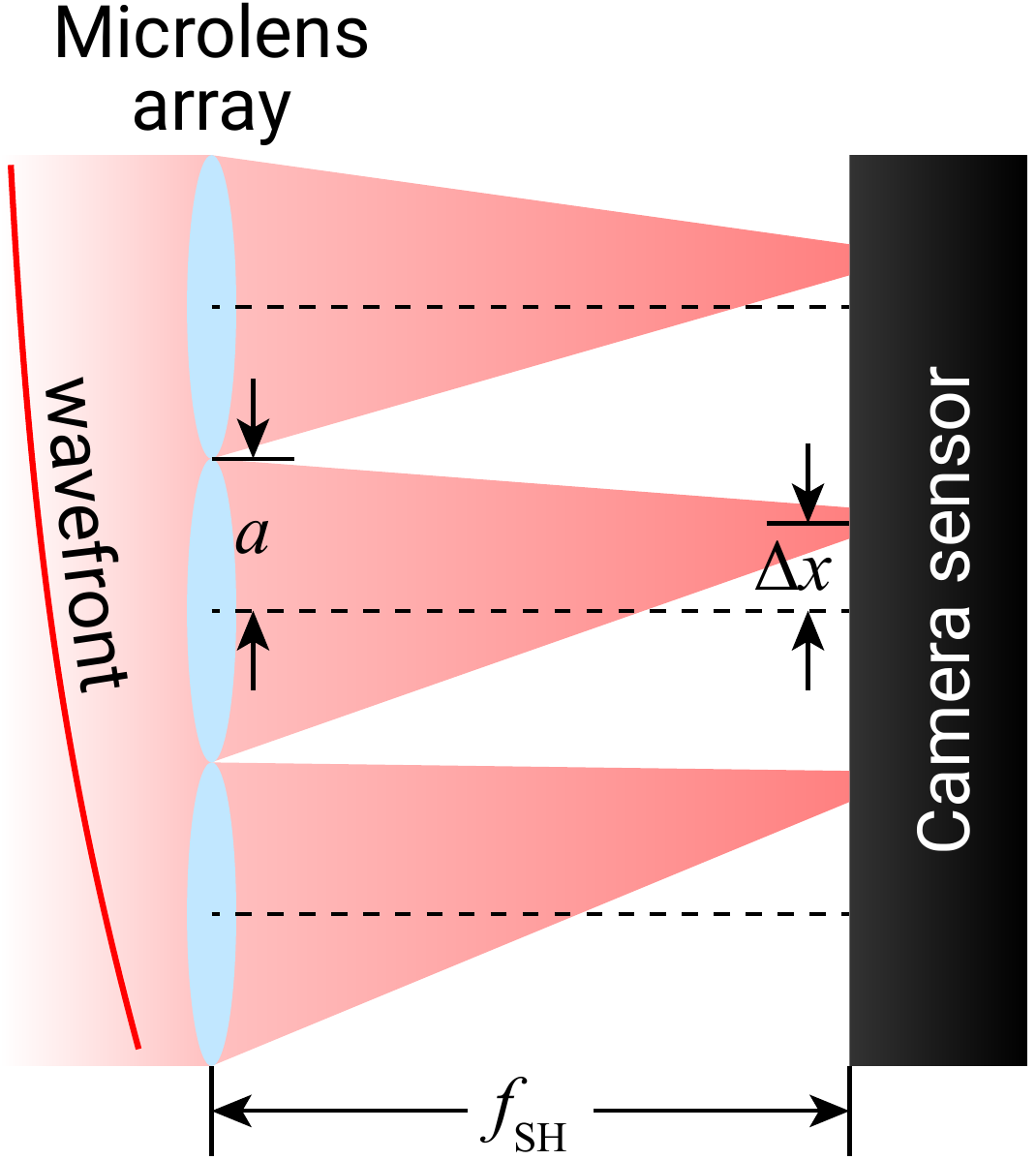}
\caption{\label{supshws}An illustration of classical Shark--Hartmann wavefront sensor.}
\end{figure*}

Here we review the basic principle and limitations of classical SHWS. The one-dimensional case is illustrated in Fig.\ \ref{supshws}. A monochromatic light beam at $\lambda$ with a curved wavefront is incident on the microlens array whose focal length is $f_{\mathrm{SH}}$. On each microlens aperture with a width $2a$, the beam can be regarded as a tilted plane wave. So, measuring the centroid position of the spot at the back focal plane of a certain microlens $\Delta x$, the tilt angle is $\arctan(\Delta x/f_{\mathrm{SH}})$, and the transverse component of the wave vector $k_x=2\pi\sin[\arctan(\Delta x/f_{\mathrm{SH}})]/\lambda$. Evaluating all the microlenses, we have a distribution of $k_x$, denoted as the function $k_x(x)$. SHWS is insensitive to polarization, and according to dispersion, $f_{\mathrm{SH}}$ has little dependence on $\lambda$, which may be negligible.

Denoting the complex amplitude of the beam as $U(x)$ and its phase $\phi(x)=\operatorname{arg}U(x)$, by analyzing the average angular spectrum of the optical field within one microlens aperture, letting $x_0$ be the center of one microlens, we have [1]
\begin{equation}\label{classicalshwsk}
    k_x(x_0)=\frac{\int_{x_0-a}^{x_0+a}dx\phi'(x)|U(x)|^2}{\int_{x_0-a}^{x_0+a}dx|U(x)|^2},
\end{equation}
which means $k_x$ is the weighted average of the phase derivative $\phi'(x)$ over the aperture according to the intensity $|U(x)|^2$. If the phase derivative does not change much inside each aperture, we have $\phi'(x)\approx k_x(x)$. Integrating $\phi'(x)$ yields the phase distribution we would like to measure. By summing the intensity values within each aperture one also obtains the intensity distribution $I(x)$, then the complex amplitude can be reconstructed $U(x)=\sqrt{I(x)}\phi(x)$. So, SHWS simultaneously measures the intensity and the phase gradient distribution at a reduced spatial resolution.

If the beam is partially coherent described by the mutual coherence function $J(x',x)$ (at the same time; corresponding to the density matrix in the position basis [2]), the phase is not defined, and the measured ``phase derivative'' using SHWS is [2]
\begin{equation}\label{mixedstategrad}
    \frac{\int_{x_0-a}^{x_0+a}dx\operatorname{Im}\partial_1J(x,x)}{\int_{x_0-a}^{x_0+a}dxJ(x,x)},
\end{equation}
which becomes Eq.\ \eqref{classicalshwsk} when the field is coherent $J(x',x)=U(x')U^\ast(x)$. Here $\partial_1$ means taking the partial derivative of the first variable, i.e., $\partial_1J(x,x)=\partial_{x'}J(x',x)|_{x'=x}$. When the aperture at $x$ is small, Eq.\ \eqref{mixedstategrad} is approximated by $[\operatorname{Im}\partial_1J(x,x)]/J(x,x)=\nabla_1\operatorname{arg}J(x,x)$. One can verify that when the partially coherent beam passes through a pure-phase object $\varphi(x)$, its mutual coherence function becomes $J(x',x)e^{i[(\varphi(x')-\varphi(x)]}$, and $\varphi'(x)$ is added to the measured ``phase derivative'' [3]. This means one does not need perfect coherent beam to illuminate a pure-phase object in order to measure its optical path difference distribution, but the focused spot size will be larger.

Eq.\ \eqref{mixedstategrad} can be generalized to the two-dimensional case. Let $S$ be the aperture with its center $(x_0,y_0)$ and width $2a$. First, we only consider the measured ``partial derivative of phase with respect to $x$'', which is calculated from the $x$ coordinate of the spot centroid. The original mutual coherence function is $J(\boldsymbol{\rho}',\boldsymbol{\rho})=J(x',y',x,y)$. Ignoring the $y$ coordinate inside the aperture, the result is similar as a partial trace $J_x(x',x)=\int_{y_0-a}^{y_0+a}dyJ(x',y,x,y)$. Substituting it into Eq.\ \eqref{mixedstategrad} yields
\begin{equation}
    \frac{\int_{x_0-a}^{x_0+a}dx\int_{y_0-a}^{y_0+a}dy\operatorname{Im}\partial_1J(x,y,x,y)}{\int_{x_0-a}^{x_0+a}dx\int_{y_0-a}^{y_0+a}dyJ(x,y,x,y)}=\frac{\int_{S}dxdy\operatorname{Im}\partial_1J(x,y,x,y)}{\int_{S}dxdyJ(x,y,x,y)}.
\end{equation}
Similarly, for ``partial derivative of phase with respect to $y$'', the result is
\begin{equation}
    \frac{\int_{S}dxdy\operatorname{Im}\partial_2J(x,y,x,y)}{\int_{S}dxdyJ(x,y,x,y)}.
\end{equation}
Combined as a vector, the measured ``phase gradient'' by SHWS at aperture $S$ from the partially coherent light $J(\boldsymbol{\rho}',\boldsymbol{\rho})$ is
\begin{equation}\label{twodmixgrad}
    \frac{\int_{S}d\boldsymbol{\rho}\operatorname{Im}\nabla_1J(\boldsymbol{\rho},\boldsymbol{\rho})}{\int_{S}d\boldsymbol{\rho}J(\boldsymbol{\rho},\boldsymbol{\rho})}.
\end{equation}

The spatial resolution determined by the microlens size is the major shortcoming of SHWS. If $U(x)$ varies rapidly in neighboring apertures (e.g., the beam width is comparable to or smaller than the aperture), SHWS cannot correctly reconstruct the phase. If sharp changes in the measured intensity distribution, or larger and distorted focused spots inside apertures are observed, the user can know the spatial resolution is not enough, and a $4f$ system should be used to magnify the beam. Also, if $U(x)$ has zeroes (e.g.\ Hermite--Gaussian beams), the phase differences between different spots cannot be detected.

If the microlens size $2a$ decreases, the spatial resolution is enhanced, but the light spots on the camera sensor becomes larger, which means the local momentum measurement is less certain. This exemplifies the uncertainty principle. However, for SHWS, only the average positions of the spots are used, so larger spots do not limit SHWS in principle, only affecting its efficiency. Another disadvantage is that the dynamic range will decrease, i.e., the maximally detectable tilt angle becomes small.

\vspace{2em}
\centerline{\bf B.\;Full derivation of Eq.\ (1) of the main text}
\vspace{1em}

Using the mixed state described by the density matrix (unnormalized) at the position basis
\begin{equation}
\varrho_2(\boldsymbol{\rho}'_2,\boldsymbol{\rho}_2)=\int_{S_1}d\boldsymbol{\rho}_1\psi(\boldsymbol{\rho}_1,\boldsymbol{\rho}'_2)\psi^\ast(\boldsymbol{\rho}_1,\boldsymbol{\rho}_2),
\end{equation}
according to Eq.\ \eqref{twodmixgrad}, the measured ``phase gradient'' by SHWS at aperture $S_2$, which is calculated from the spot centroid position, is
\begin{align}
\frac{\int_{S_2}d\boldsymbol{\rho}_2\operatorname{Im}\nabla_1\varrho_2(\boldsymbol{\rho}_2,\boldsymbol{\rho}_2)}{\int_{S_2}d\boldsymbol{\rho}_2\varrho_2(\boldsymbol{\rho}_2,\boldsymbol{\rho}_2)}
&=\frac{\int_{S_1}d\boldsymbol{\rho}_1\int_{S_2}d\boldsymbol{\rho}_2\operatorname{Im}\nabla_{\boldsymbol{\rho}'_2}[\psi(\boldsymbol{\rho}_1,\boldsymbol{\rho}'_2)\psi^\ast(\boldsymbol{\rho}_1,\boldsymbol{\rho}_2)]|_{\boldsymbol{\rho}'_2=\boldsymbol{\rho}_2}}{\int_{S_1}d\boldsymbol{\rho}_1\int_{S_2}d\boldsymbol{\rho}_2\psi(\boldsymbol{\rho}_1,\boldsymbol{\rho}_2)\psi^\ast(\boldsymbol{\rho}_1,\boldsymbol{\rho}_2)}\nonumber\\
&=\frac{\int_{S_1}d\boldsymbol{\rho}_1\int_{S_2}d\boldsymbol{\rho}_2\operatorname{Im}\{[\nabla_2\psi(\boldsymbol{\rho}_1,\boldsymbol{\rho}_2)]\psi^\ast(\boldsymbol{\rho}_1,\boldsymbol{\rho}_2)\}}{\int_{S_1}d\boldsymbol{\rho}_1\int_{S_2}d\boldsymbol{\rho}_2|\psi(\boldsymbol{\rho}_1,\boldsymbol{\rho}_2)|^2}\nonumber\\
&=\frac{\int_{S_1}d\boldsymbol{\rho}_1\int_{S_2}d\boldsymbol{\rho}_2|\psi(\boldsymbol{\rho}_1,\boldsymbol{\rho}_2)|^2\operatorname{Im}\frac{\nabla_2\psi(\boldsymbol{\rho}_1,\boldsymbol{\rho}_2)}{\psi(\boldsymbol{\rho}_1,\boldsymbol{\rho}_2)}}{\int_{S_1}d\boldsymbol{\rho}_1\int_{S_2}d\boldsymbol{\rho}_2|\psi(\boldsymbol{\rho}_1,\boldsymbol{\rho}_2)|^2}\nonumber\\
&=\frac{\int_{S_1}d\boldsymbol{\rho}_1\int_{S_2}d\boldsymbol{\rho}_2|\psi(\boldsymbol{\rho}_1,\boldsymbol{\rho}_2)|^2\nabla_2\phi(\boldsymbol{\rho}_1,\boldsymbol{\rho}_2)}{\int_{S_1}d\boldsymbol{\rho}_1\int_{S_2}d\boldsymbol{\rho}_2|\psi(\boldsymbol{\rho}_1,\boldsymbol{\rho}_2)|^2},
\end{align}
where $\operatorname{Im}\frac{\nabla_2\psi(\boldsymbol{\rho}_1,\boldsymbol{\rho}_2)}{\psi(\boldsymbol{\rho}_1,\boldsymbol{\rho}_2)}=\operatorname{Im}\nabla_2\ln\psi(\boldsymbol{\rho}_1,\boldsymbol{\rho}_2)=\nabla_2\operatorname{arg}\psi(\boldsymbol{\rho}_1,\boldsymbol{\rho}_2)=\nabla_2\phi(\boldsymbol{\rho}_1,\boldsymbol{\rho}_2)$.
\end{spacing}
\begin{spacing}{1.12}
\vspace{2em}
\centerline{\bf C.\;Details of the experimental setup}
\vspace{1em}

A laser with $\lambda_p=405~\textrm{nm}$ passes through a beam shaping system to ensure that the beam is roughly Gaussian, and pumps the BBO crystal with a thickness $L=5~\textrm{mm}$. The pump beam at the crystal has a beam radius of about $0.8$~mm, and is then removed by a long-pass interference filter (IF) at $473$~nm. Two Fourier lenses with $f_1=5$~cm and $f_2=30$~cm magnify the beam $6$ times to the SLM (Hamamatsu LCOS-SLM X13138-02. Pixel size: $12.5$~\textmu m). The light reflected by the SLM passes through a third Fourier lens with $f_3=50$~cm and is incident on the microlens array with the lens width $0.295$~mm, the array period $0.3$~mm, the focal length $14.6$~mm and the total size $1~\mathrm{cm}\times1~\mathrm{cm}$. Since the effective clear aperture of the microlens array is $9$~mm $\times$ $9$~mm, the number of effective microlenses is $30\times30$. As the focal length of the microlens array is less than the distance between the EMCCD (Andor iXon Ultra 897 with $512\times512$ pixels; Pixel width: $16$~\textmu m) sensor and the camera casing, an imaging lens ($f_4=4$~cm) is placed at the midpoint of the focal plane of the microlenses and the sensor so that the light field at the focal plane is imaged onto the sensor with a magnification factor $-1$. Installing the microlens array directly before the sensor is recommended to avoid the aberration of the imaging lens. A bandpass IF [$(810\pm5)$\,nm, Thorlabs] is inserted before the sensor to approximately select degenerate down-converted photons with $\lambda=810$~nm. The microlens array, the imaging lens and the bandpass IF are installed on an optical cage system connected to the EMCCD so that they can be moved together. During the experiments, the EMCCD sensor is cooled to $-60^\circ\mathrm{C}$. The exposure time, EM gain, horizontal pixel readout rate, vertical
pixel shift speed and vertical clock voltage amplitude are set to $1$~ms, $1000$, $17$~MHz, $0.3$~\textmu s and $+4$~V respectively. A region of interest (ROI) is selected for each experiment when taking frames (See Section G).

\vspace{2em}
\centerline{\bf D.\;Multiple frame method, brightness separation formula and comparison}
\vspace{1em}

We mentioned the original covariance formula [4,5] and the successive frame formula [4,6,7] of the multiple frame method in the main text. Here we briefly describe this method. After thresholding, we denote the value ($0$ or $1$) at the $i$th pixel (The pixel indexing is arbitrary) of the $n$th frame as $C_{n,i}$, and the discretized JPD of the $i$th and $j$th pixel as $\Gamma_{ij}$. Defienne \emph{et al.}\ showed that the JPD is approximately proportional to the covariance of the counting at the two pixels $\Gamma_{ij}\approx\operatorname{Cov}_{ij}=\langle C_{ij}\rangle-\langle C_{i}\rangle\langle C_{j}\rangle$, where $\langle C_{i}\rangle=\sum_{n=1}^NC_{n,i}/N$ is the average count of the $i$th pixel and $\langle C_{ij}\rangle=\sum_{n=1}^NC_{n,i}C_{n,j}/N$ is the average product of the counts at the $i$th and $j$th pixel, which we name the covariance formula. Its derivation has considered the quantum efficiency and dark counts, and assumes Poissonian statistics of photons, while in experiments the beam intensity and the sensor efficiency are affected by many factors. If either of them fluctuates with time, $\langle C_{ij}\rangle$ becomes larger than $\langle C_{i}\rangle\langle C_{j}\rangle$ when the actual $\Gamma_{ij}=0$, leading to a positive background in the calculated JPD. The successive frame formula is that the term $\langle C_{i}\rangle\langle C_{j}\rangle$ is replaced by $\sum_{n=1}^NC_{n,i}C_{n+1,j}/N$, which may overcome this problem [4], but does not succeed here.

\begin{figure*}[h!]
\includegraphics[width=0.98\textwidth]{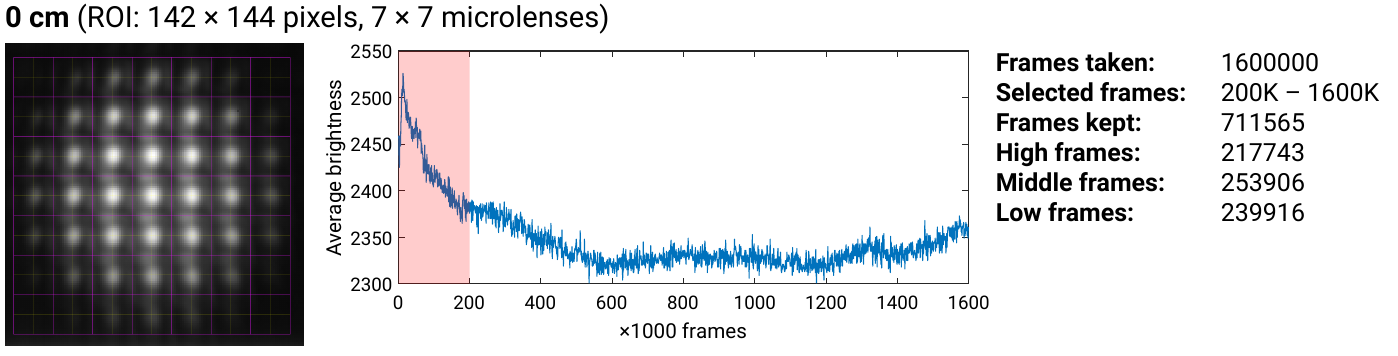}
\includegraphics[width=0.98\textwidth]{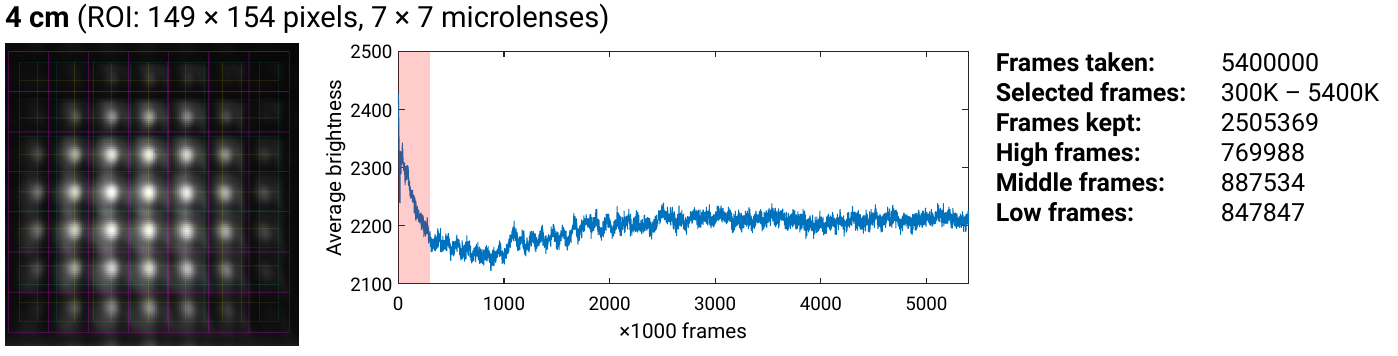}
\includegraphics[width=0.98\textwidth]{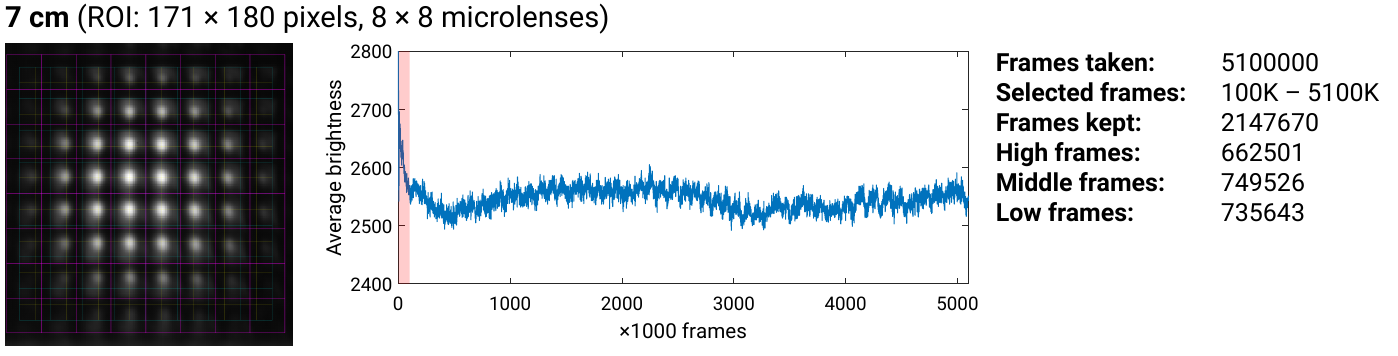}
\includegraphics[width=0.98\textwidth]{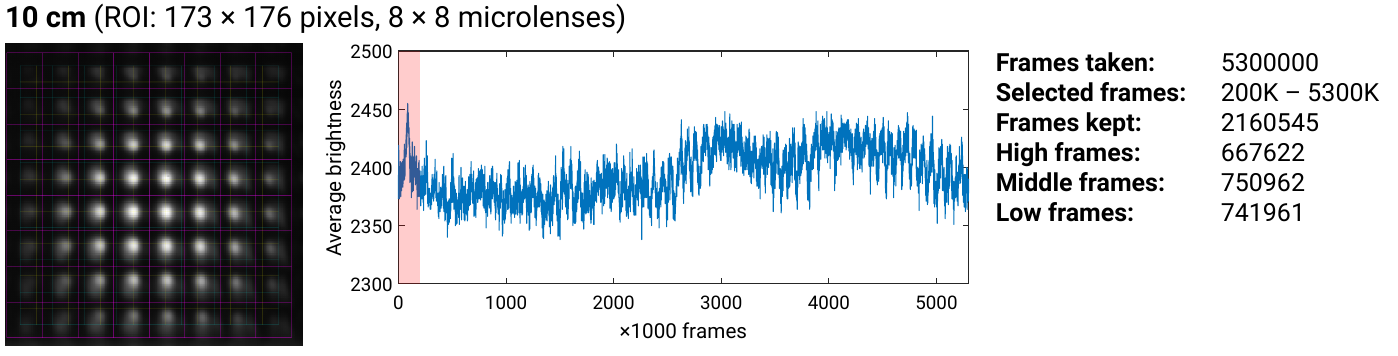}
\includegraphics[width=0.98\textwidth]{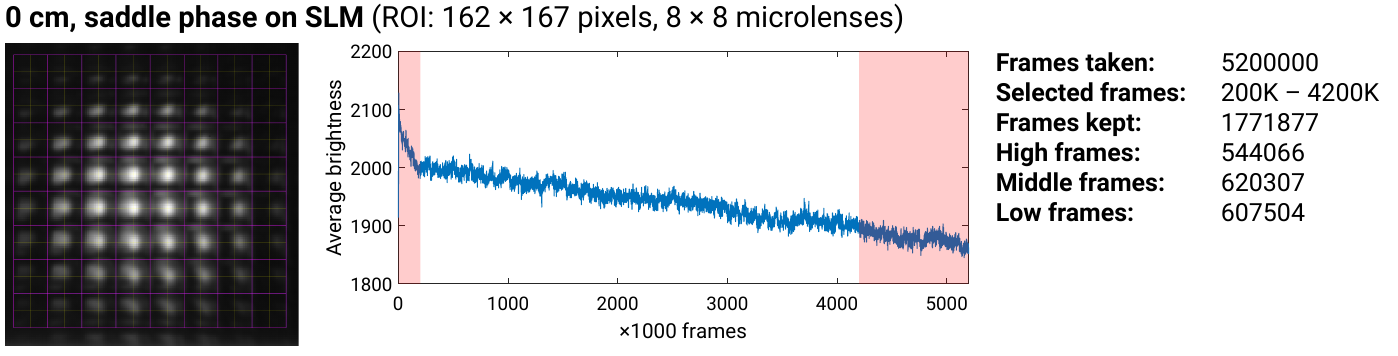}
\caption{\label{supdetails}The direct image, a graph of average brightness values per $1000$ frames, and related data (``K'' means $\times1000$) of each experiment.}
\end{figure*}

In our experiments, the thresholding process is that pixels whose grayscale values (from $0$ to $65535$) are greater than $215$ are recorded as $C_{n,i}=1$. The top $2$ lines in the frames are cropped to avoid abnormal dark counts caused by the EMCCD, and the images are then rotated by $180^\circ$ to undo the effect of the imaging lens. We define the brightness of a frame as the number of pixels whose $C_{n,i}=1$. A graph of the average brightness per $1000$ frames is plotted, and a subset from the original image data is selected to ensure the variation range of the average brightness per $1000$ frames is about (or less than) $100$ (Fig.\ \ref{supdetails} provides the brightness graph and the subset selection of each experiment). In our brightness separation formula, the average brightness $\bar{B}$ of frames of the subset is calculated. Then, one frame is named:

(1) low frame if its brightness $B$ satisfies $\bar{B}-3\sqrt{\bar{B}}\leq B<\bar{B}-\sqrt{\bar{B}}$;

(2) middle frame if $\bar{B}-\sqrt{\bar{B}}\leq B<\bar{B}+\sqrt{\bar{B}}$;

(3) high frame if $\bar{B}+\sqrt{\bar{B}}\leq B<\bar{B}+3\sqrt{\bar{B}}$;

(4) otherwise it is discarded.

Letting $N_\mathrm{L},N_\mathrm{M},N_\mathrm{H}$ be the number of low, middle and high frames respectively, and $S_\textrm{L,$i$}$ be $\sum_\textrm{low $n$}C_{n,i}$ (similar expressions for $S_\textrm{M,$i$}$ and $S_\textrm{H,$i$}$), the JPD is calculated by
\begin{equation}
    \Gamma_{ij}=\left(\sum_\textrm{low, middle and high $n$}C_{n,i}C_{n,j}\right)-\frac{S_\textrm{L,$i$}S_\textrm{L,$j$}}{N_\mathrm{L}}-\frac{S_\textrm{M,$i$}S_\textrm{M,$j$}}{N_\mathrm{M}}-\frac{S_\textrm{H,$i$}S_\textrm{H,$j$}}{N_\mathrm{H}},
\end{equation}
which can be understood as $N_{\textrm{L}}\operatorname{Cov}_{\textrm{L},ij}+N_{\textrm{M}}\operatorname{Cov}_{\textrm{M},ij}+N_{\textrm{H}}\operatorname{Cov}_{\textrm{H},ij}$, where $\operatorname{Cov}_{\mathrm{L},ij},\operatorname{Cov}_{\mathrm{M},ij},\operatorname{Cov}_{\mathrm{H},ij}$ are from the covariance formula of low, middle, high frames respectively. Calculating respectively can avoid the background from the brightness fluctuation. Weighting according to $N_\mathrm{L},N_\mathrm{M},N_\mathrm{H}$ reduces the impact of covariance error from a lower number of frames (As can be seen in Fig.\ \ref{supdetails}, $N_{\mathrm{L}},N_{\mathrm{M}},N_{\mathrm{H}}$ are approximately equal for each experiment, so the impact is not significant). If all frames are of one type (e.g.\ middle) and not discarded (defining $N_{\textrm{L}}\operatorname{Cov}_{\textrm{L},ij}=N_{\textrm{H}}\operatorname{Cov}_{\textrm{H},ij}=0$ when $N_{\textrm{L}}=N_{\textrm{H}}=0$), this formula reverts to the original covariance formula $\langle C_{ij}\rangle-\langle C_{i}\rangle \langle C_{j}\rangle$ times the number of frames. 

The JPD of the same pixel $\Gamma_{ii}$ cannot be measured in principle [6], and counts of two pixels on the same line have abnormal correlation (The closer they are, the stronger the correlation is) from the mechanism of EMCCD, so we use interpolation to eliminate these effects. If the two pixels $(x_1,y_1)$ and $(x_2,y_2)$ satisfy $y_1=y_2$ and $|x_1-x_2|\leq8$, $\Gamma_{x_1,y_1,x_2,y_2}$ is replaced by $(\Gamma_{x_1,y_1,x_2,y_2+1}+\Gamma_{x_1,y_1,x_2,y_2-1})/2$ (if $y_2$ takes the top or bottom value, it is $\Gamma_{x_1,y_1,x_2,y_2+1}$ or $\Gamma_{x_1,y_1,x_2,y_2-1}$ respectively).

\begin{figure*}[h]
\includegraphics[width=0.65\textwidth]{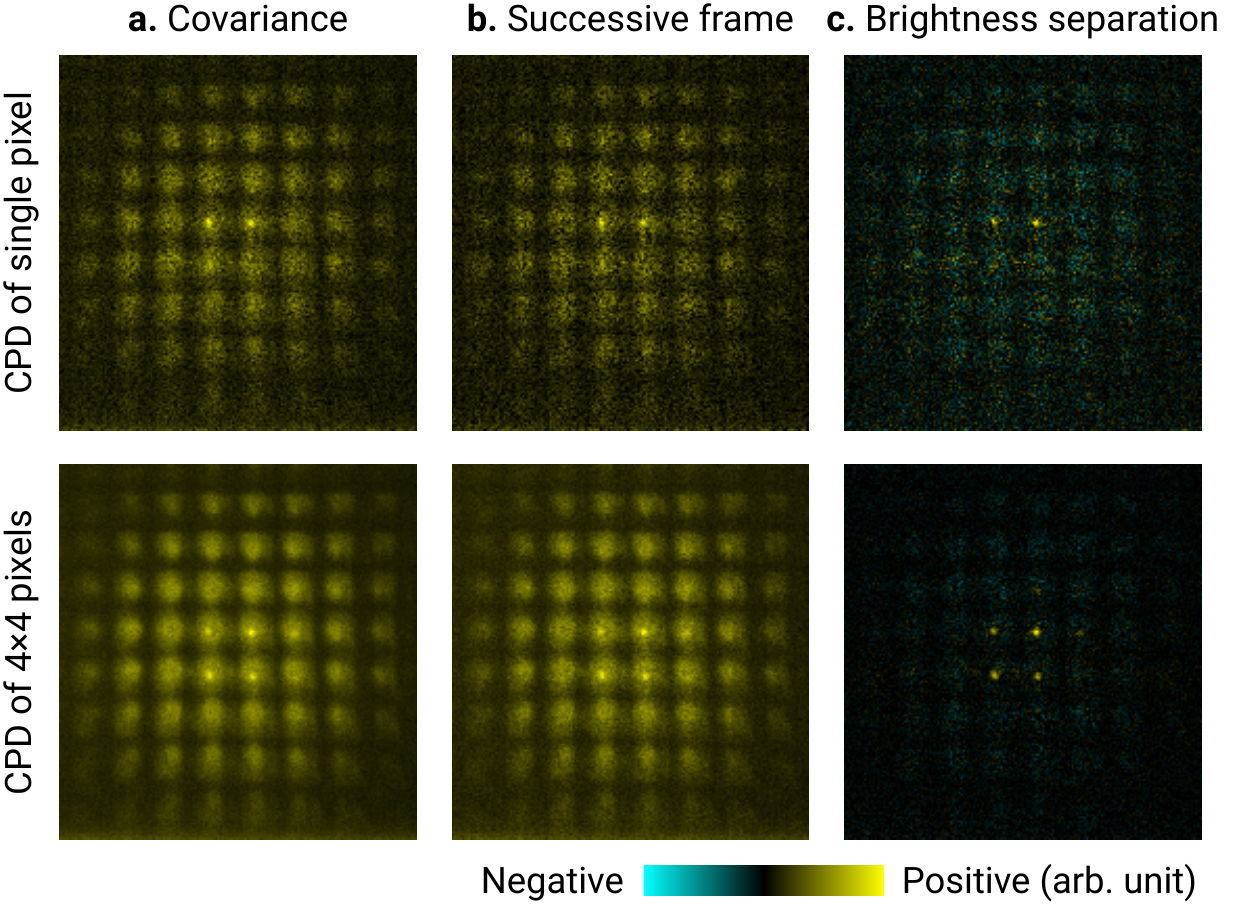}
\caption{\label{supcompare}Comparison of different JPD calculation formulas. The single-pixel CPDs and their sums over $4\times4$ given pixels (The same location as Fig.\ 2(b) of the main text) calculated from the $7$-cm propagation data are shown.}
\end{figure*}

Here we also use the other two formulas to calculate the single-pixel CPDs and their sums over $4\times4$ given pixels as shown in Fig.\ \ref{supcompare}, and compare them with our formula. We find that the bright spots are also visible, but the background noises are significant. Using the brightness separation formula, the JPD values in regions with no coincidence are mainly negative, possibly because the sub-Poissonian photon statistics after frame discarding introduces a slightly negative background.

The methods and algorithms used in our experiments are not perfect, and can be improved in the future.

\vspace{2em}
\centerline{\bf E.\;Details of the calculation of the aperture CPD and the phase gradient}
\vspace{1em}

\begin{figure*}[h]
\includegraphics[width=0.38\textwidth]{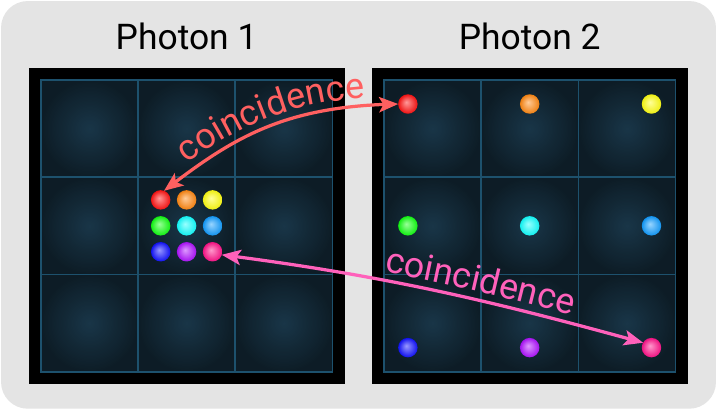}
\caption{\label{supprinillfig}An illustration of the relation between coincident photons, single-pixel CPD and aperture CPD. Different colors represent different coincident photon pairs, and the distinguishable case is considered here. When photon 1 is postselected to different positions within a microlens aperture, photon 2 is projected at different positions inside different apertures, i.e., the single-pixel CPD of photon 2 covers different regions when the given points (the positions of photon 1) are different. Summing the single-pixel CPDs of photon 2 with the given point taking all pixels within an aperture yields the aperture CPD.}
\end{figure*}

In this section, we describe the data processing from single-pixel CPDs to aperture CPDs. An illustration of its principle is shown in Fig.\ \ref{supprinillfig}. First, we evaluate the direct image (the sum of all the thresholded frames) and determine the position of the apertures (See Fig.\ \ref{supdetails}). The apertures used in our calculations are not necessarily the same as the microlenses. In the SPDC biphoton propagation data, the apertures used in data processing are larger to contain the whole spot. 

\begin{figure*}[h]
\includegraphics[width=0.98\textwidth]{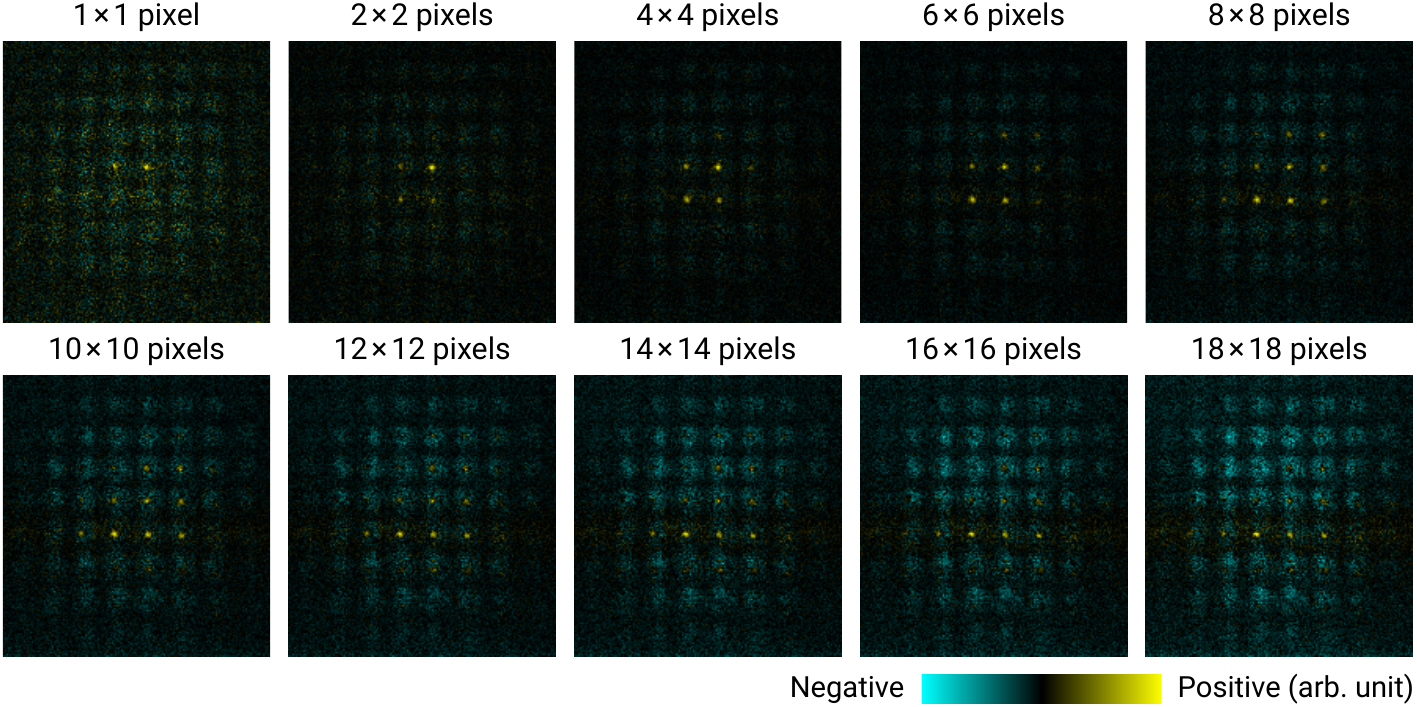}
\caption{\label{sup18}Direct summation of single-pixel CPDs over regions of different sizes centered by the same pixel from the $7$-cm data.}
\end{figure*}

All pixels completely inside each aperture are considered in the calculation. However, if the CPDs of them are directly summed, the result is distorted (Lines where the given points lie are brighter than other lines) due to the abnormal correlation of EMCCD. Fig.\ \ref{sup18} shows direct summations of single-pixel CPDs over regions with different sizes, ranging from the original single-pixel CPD ($1\times1$ pixel) to the sum over almost the whole aperture ($18\times18$ pixels). We find that summing over $4\times4$ pixels has a better performance.

\begin{figure*}[h!]
\includegraphics[width=\textwidth]{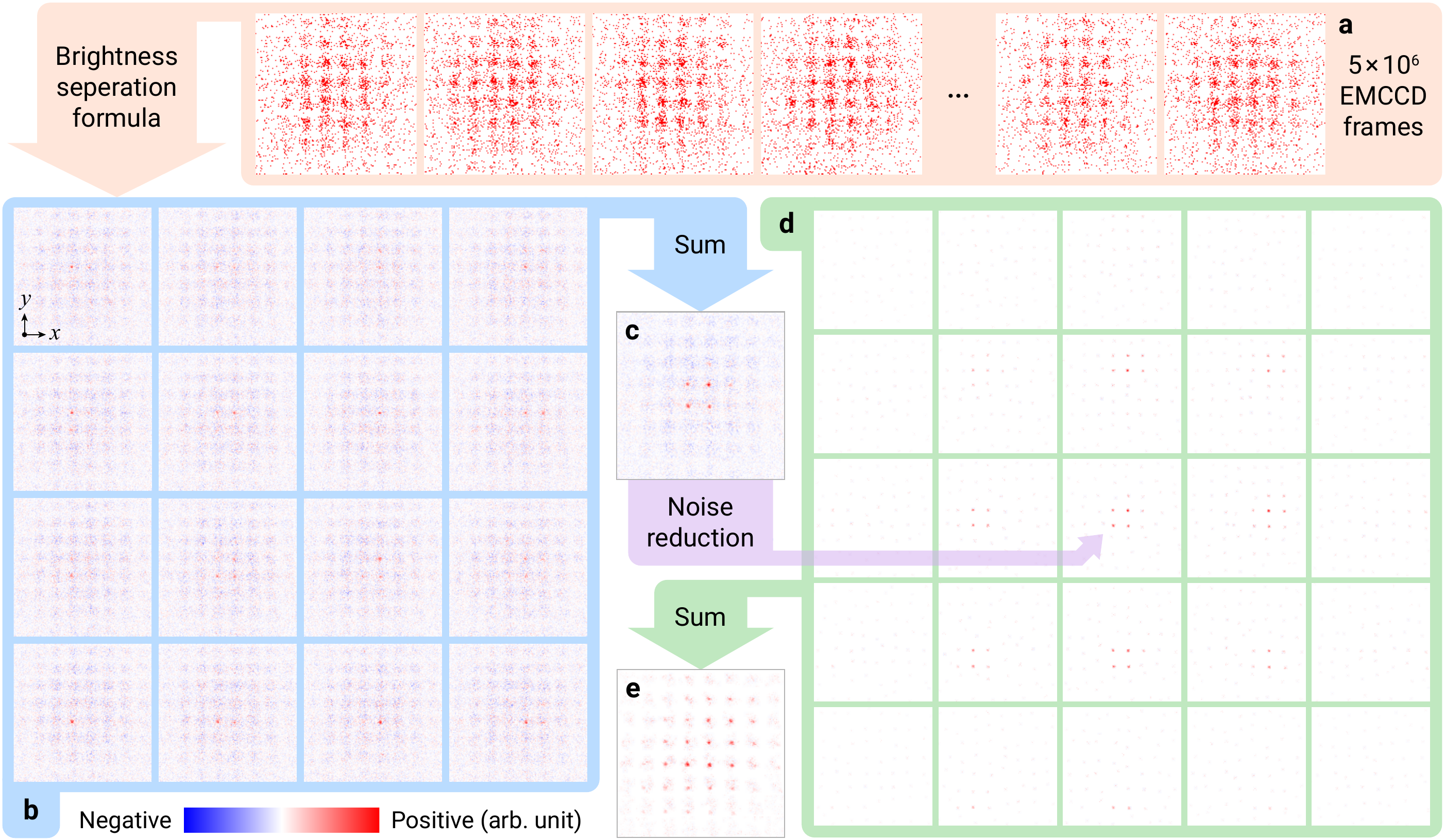}
\caption{\label{supprinfig}A more complete illustration of the calculation of aperture CPD. \textbf{a}, Thresholded EMCCD frames. \textbf{b}, Single-pixel CPDs of $4\times4$ pixels in a segment. \textbf{c}, The sum of $4\times4$ CPDs in \textbf{b}. \textbf{d}, The CPDs of all the $4\times4$-pixel segments inside an aperture after noise reduction. \textbf{e}, The sum of all the distributions in \textbf{d}. The aperture CPD is obtained by setting the negative values in \textbf{e} to zero.}
\end{figure*}

\begin{figure*}[h!]
\includegraphics[width=0.87\textwidth]{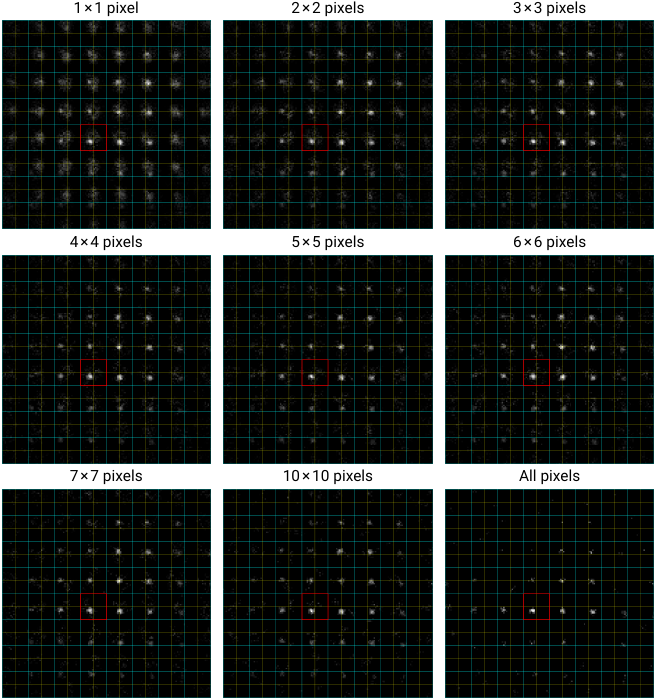}
\caption{\label{supaptcpd}The calculated aperture CPDs after initial summing with different segment sizes, denoising and final summing from the $7$-cm data. The cyan grid indicates the microlens borders, and the yellow grid indicates the microlens centers. The given aperture is marked by red boxes.}
\end{figure*}

In our data processing, each aperture is divided into $4\times4$-pixel segments (segments near the right and top border of the aperture may be smaller), and the $16$ (or fewer) CPDs of each segment are summed first (as shown in the process from Fig.\ \ref{supprinfig}b to \ref{supprinfig}c). The result goes through the noise reduction process. In each aperture, the maximum value is found and all negative values are set to zero. The values inside a $5\times5$ region centered by the maximum pixel are subtracted by the average value of pixels adjacent to the $5\times5$ region (i.e., inside a $7\times7$ region centered by the maximum pixel excluding the $5\times5$ region), so that the background is subtracted (negative values can be introduced again). Finally, pixels outside the $5\times5$ region are set to zero.

After noise reduction, the CPDs of all the $4\times4$-pixel segments are summed, as shown in Fig.\ \ref{supprinfig}e, and negative values of the sum are set to zero. This becomes the final aperture CPD. Fig.\ \ref{supaptcpd} also provides the calculated aperture CPDs with various segment sizes. Sizes ranging from $3\times3$ to $7\times7$ pixels can all yield a satisfactory result with less noise or amplitude distortion.

When calculating the phase gradient distribution from an aperture CPD, a $3\times3$-pixel region with the largest CPD sum is chosen, and the $7\times7$ region centered by the $3\times3$ region is used to calculate the centroid position of the spot $\Delta\boldsymbol{\rho}$ relative to the microlens center (The rest of the aperture is believed to have no light spot, and does not contribute to the calculation).

\vspace{2em}
\centerline{\bf F.\;The phase reconstruction algorithm}
\vspace{1em}

With the phase gradient distributions $\boldsymbol{k}_1,\boldsymbol{k}_2$, to realize the line integral in Eq.\ (2) of the main text, we use the computer algorithm based on random point spreading and averaging in our previous work [3], whose steps are as follows:

(1) Select the point with the maximum JPD $\Gamma_{\mathrm{max}}$ as the starting point whose phase is $0$. A thread-safe queue is used and the point is enqueued. The phase values of all other points are marked as ``not calculated''. Mark the status of all threads as ``unfinished'' and start multiple threads ($10$ threads in our experiments).

(2) A single thread attempts to dequeue a point. If the queue is empty, go to step (3). Otherwise, the thread marks its status as ``unfinished'', checks how many adjacent points (up to $8$ points) of the dequeued one are not calculated, and then probabilistically chooses whether to calculate the phase of the adjacent point based on the zonal method formula and enqueue it, or to ignore it. Denoting the coordinates of the dequeued point as $(m_1,n_1,m_2,n_2)$ ($m_1,n_1,m_2,n_2$ are integers), the basic expansion formulas are, for example,
\begin{align}
	&\phi(m_1-1,n_1,m_2,n_2)=\phi(m_1,n_1,m_2,n_2)-k_{1x}(m_1-1,n_1,m_2,n_2)d,\nonumber\\
	&\phi(m_1+1,n_1,m_2,n_2)=\phi(m_1,n_1,m_2,n_2)+k_{1x}(m_1,n_1,m_2,n_2)d,\nonumber\\
	&\phi(m_1,n_1,m_2-1,n_2)=\phi(m_1,n_1,m_2,n_2)-k_{2x}(m_1,n_1,m_2-1,n_2)d,\nonumber\\
	&\phi(m_1,n_1,m_2+1,n_2)=\phi(m_1,n_1,m_2,n_2)+k_{2x}(m_1,n_1,m_2,n_2)d,
\end{align}
where $d$ is the spacing of the discretized gradient distributions (i.e., microlens aperture width divided by the interpolation times). If the JPD of the dequeued point $\Gamma$ is smaller, the probability of calculation $\Gamma/(2\Gamma_{\mathrm{max}})+0.1$ is also smaller due to larger errors. After handling all the adjacent points, if any adjacent points are ignored (still not calculated), the dequeued point is enqueued again. This thread repeats this step.

(3) This thread marks the thread status as ``finished'', checks the status of all other threads. If there are ``unfinished'' threads, go back to step (2).

(4) One phase reconstruction process is done. Record the phase distribution.

(5) To reduce errors, repeat the above process several times ($50$ times in our experiments) and take the average phase of each time as the reconstruction result.

~

\vspace{2em}
\centerline{\bf G.\;Details of the experimental data}
\vspace{1em}

Fig.\ \ref{supdetails} shows the direct images from the EMCCD, ROI sizes, the average brightness values per $1000$ frames, frame counts of each experiment. Inside the direct images, yellow grids indicate the microlens centers, cyan grids indicate the microlens borders, and magenta grids indicate the actual aperture division in data processing (If the actual aperture size equals the microlens size, the magenta grid covers the cyan one). Only a part of the frames is selected to ensure the brightness fluctuation is within $100$, and frames discarded manually are painted as light red in the graphs.

\clearpage
\end{spacing}
\begin{spacing}{1.1}
\vspace{2em}
\centerline{\bf H.\;Theory of degenerate collinear type-I SPDC and double-Gaussian parameters}
\vspace{1em}

From the phase matching condition of $\boldsymbol{k}=(\boldsymbol{q},k_z)$, assuming the pump beam is monochromatic and the transverse size of the BBO crystal is large enough so that the transverse phase matching is precise, the joint angular spectrum (wave function in the momentum space) of the degenerate down-converted photons is
\begin{equation}
    \tilde{\psi}(\boldsymbol{q}_s,\boldsymbol{q}_i)=A_p(\boldsymbol{q}_s+\boldsymbol{q}_i)\operatorname{sinc}\left[\frac{L}{2}(k_{pz}-k_{sz}-k_{iz})\right],
\end{equation}
where the symbols $p$, $s$ and $i$ stand for pump ($e$ light), signal ($o$ light) and idler ($o$ light) photon respectively, $A_p(\boldsymbol{q})$ is the angular spectrum of the pump beam, and $L$ is the crystal thickness. From the birefringent theory, 
\begin{gather}
	k_{oz}=\sqrt{\left(\frac{n_o\omega}{c}\right)^2-q_x^2-q_y^2}\approx\frac{n_o\omega}{c}-\frac{c}{2n_o\omega}(q_x^2+q_y^2),\\
	k_{ze}=\alpha q_x+\sqrt{\left(\frac{\eta\omega}{c}\right)^2-\beta^2q_x^2-\gamma^2q_y^2}\approx\alpha q_x+\frac{\eta\omega}{c}-\frac{c}{2\eta\omega}(\beta^2q_x^2+\gamma^2q_y^2),
\end{gather}
where $\omega$ is the frequency (invariant in media) of the light considered, $c$ is the speed of light in vacuum,
\begin{equation}
	\alpha=\frac{(n_o^2-n_e^2)\sin\theta\cos\theta}{n_o^2\sin^2\theta+n_e^2\cos^2\theta},\eta=\frac{n_on_e}{\sqrt{n_o^2\sin^2\theta+n_e^2\cos^2\theta}},\beta=\frac{\eta^2}{n_on_e}\approx1,\gamma=\frac{\eta}{n_e}\approx1,
\end{equation}
and $\theta$ is the angle between the optical axis of the crystal (in the $x$-$z$ plane) and the surface normal line [8]. The Sellmeier equations of $n_o$ and $n_e$ of BBO can be found in Ref.\ [9]. For degenerate collinear type-I SPDC with the pump beam at $405~\mathrm{nm}$, $\theta\approx28.67^\circ$ and $\eta\approx1.6611$.

Denoting the wave number in vacuum as $k_\mathrm{vac}=\omega/c$ ($k_{\mathrm{vac},p}=2k_{\mathrm{vac}_s}=2k_{\mathrm{vac},i}$), considering the collinear phase matching, and ignoring the walk-off effect determined by $\alpha$, when $\boldsymbol{q}_s=\boldsymbol{q}_i=\boldsymbol{0}$, we have $\eta_p=n_{so}=n_{io}$ and
\begin{equation}
    k_{pz}-k_{sz}-k_{iz}=\alpha_p(q_{sx}+q_{ix})+\frac{|\boldsymbol{q}_s-\boldsymbol{q}_i|^2}{2\eta_pk_{\mathrm{vac},p}}\approx\frac{|\boldsymbol{q}_s-\boldsymbol{q}_i|^2}{2k_p},
\end{equation}
\begin{equation}
    \tilde{\psi}(\boldsymbol{q}_s,\boldsymbol{q}_i)\approx A_p(\boldsymbol{q}_s+\boldsymbol{q}_i)\operatorname{sinc}\left(\frac{L|\boldsymbol{q}_s-\boldsymbol{q}_i|^2}{4k_p}\right).
\end{equation}
If the pump beam is Gaussian with its waist at the crystal $U_p(\boldsymbol{\rho})=\exp(-|\boldsymbol{\rho}|^2/\sigma_+^2)$, its angular spectrum is $A_p(\boldsymbol{q})=\exp(-\sigma_+^2|\boldsymbol{q}|^2/4)$. However, the $\operatorname{sinc}(ax^2)$ term hinders further analytical expressions, and should be approximated by a Gaussian function. We follow the method by Chan \emph{et al.}\ that when $\operatorname{sinc}(ax^2)=1/e$, $\exp(-0.455ax^2)\approx1/e$ [10]. Letting $\sigma_-=\sqrt{0.455L/k_p}$, the wave function takes the double-Gaussian form in Eqs.\ (3) and (4) of the main text.

\begin{figure*}[t]
\includegraphics[width=\textwidth]{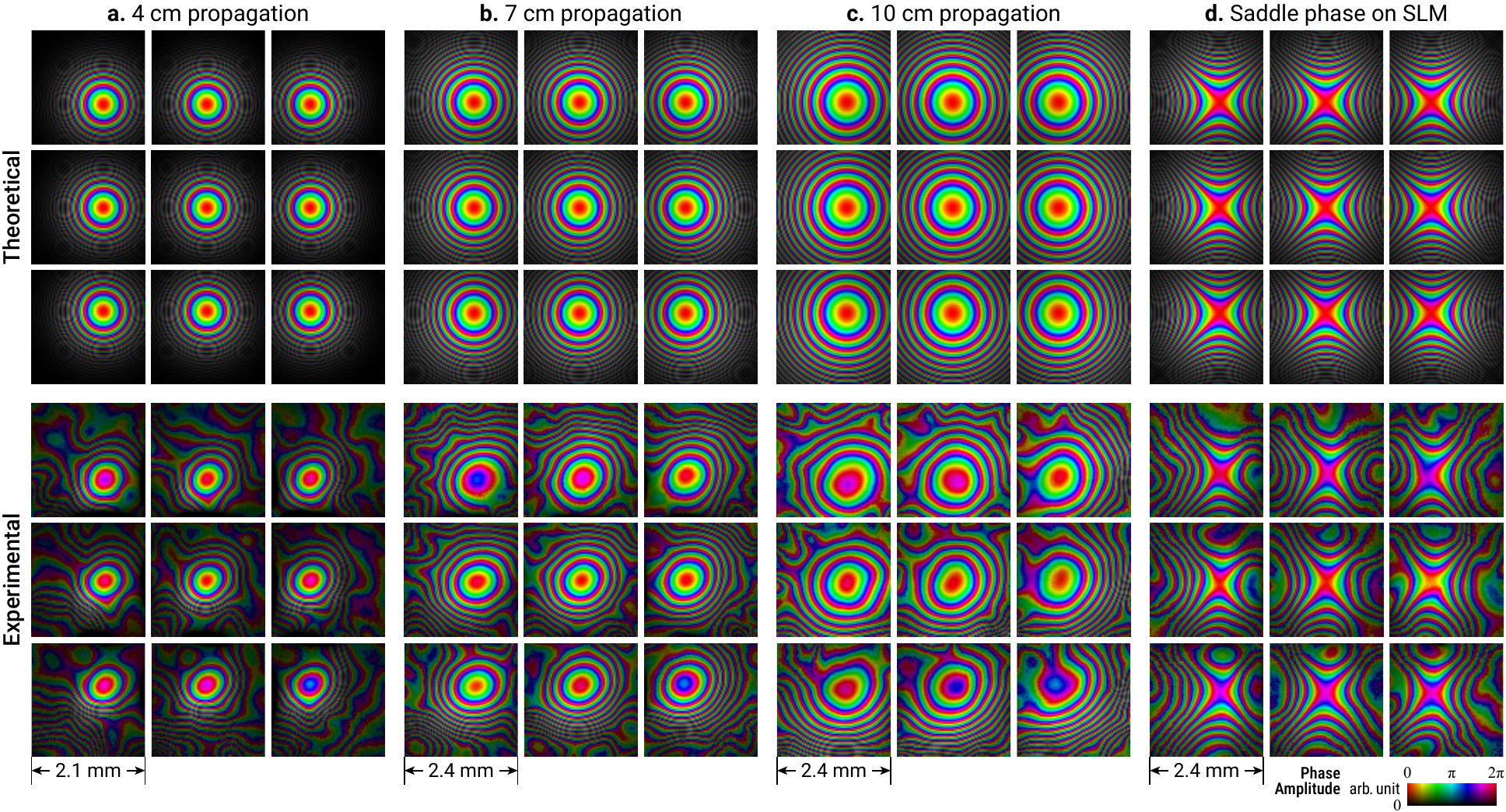}
\caption{\label{diffractsupfig}The theoretical conditional wave functions calculated from double-Gaussian functions and the experimentally reconstructed wave functions after gradient distribution interpolation of: \textbf{a}, $4$~cm propagation; \textbf{b}, $7$~cm; \textbf{c}, $10$~cm; \textbf{d}, $0$~cm with a saddle phase on SLM. The given points of the conditional wave functions are on a $3\times3$ grid with spacing $0.3$~mm, the center of which is the beam center. The global phase of each wave function is fixed when drawing the $3\times3$ conditional wave functions.}
\end{figure*}
In our experiment, the double-Gaussian parameters of photon pairs at the BBO are $\sigma_+\approx0.8$ mm and $\sigma_-\approx\sqrt{0.455L/k_p}\approx9.396$~\textmu m. The three Fourier lenses are equivalent to one lens with the focal length $f_\mathrm{eff}=25/3$~cm. $\sigma_+$ and $\sigma_-$ at its focal plane (the $0$-cm position of the microlens array) are approximately $13.43$~\textmu m and $1.143$~mm respectively, which are used to calculate the theoretical wave function after propagating a distance. The theoretical distance with no amplitude correlation is $11.91$~cm.

\vspace{2em}
\centerline{\bf I.\;Theoretical and experimental wave functions and their Fourier transforms}
\vspace{1em}

Fig.\ \ref{diffractsupfig} provides nine conditional wave functions for each theoretical and experimental case except the $0$-cm one (with no propagation). The phase pattern of conditional wave function anticorrelates with the given point.

\begin{figure*}[h]
\includegraphics[width=0.7\textwidth]{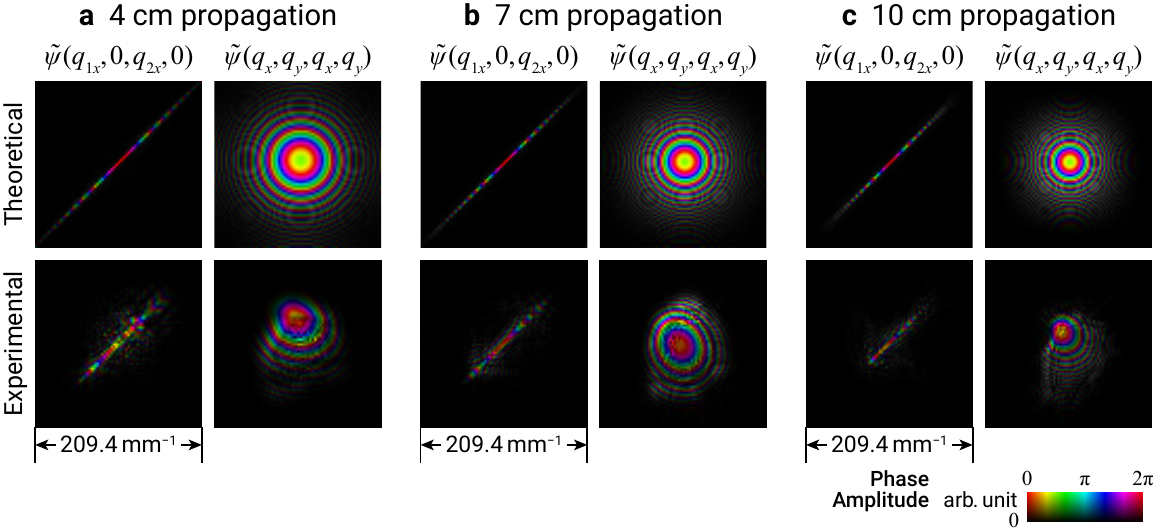}
\caption{\label{supftfig}The theoretical and experimental wave functions after Fourier transform of: \textbf{a}, $4$~cm propagation; \textbf{b}, $7$~cm; \textbf{c}, $10$~cm. The first column shows the momentum positive correlation, and the second column shows the global paraboloid phase from the angular spectrum theory.}
\end{figure*}

We perform discrete Fourier transform on the theoretical and experimental wave functions to show the biphoton wave function in momentum space. $\boldsymbol{q}_1$ and $\boldsymbol{q}_2$ are positively correlated, and the global paraboloid phase from the angular spectrum theory becomes steeper as $z$ increases. Fig.\ \ref{supftfig} shows two slices of each momentum wave function, $\tilde{\psi}(q_{1x},0,q_{2x,0})$ and $\tilde{\psi}(q_{x},q_{y},q_{x},q_{y})$. Errors of the JPD and phase make their Fourier transforms distorted, but the momentum correlation and the global phase pattern can still be seen.

~

\vspace{2em}
\centerline{\bf J.\;Analysis of classically correlated two-photon states}
\vspace{1em}

Ignoring normalization, we consider this type of classically correlated state whose density matrix is
\begin{equation}\label{classiform}
    \hat{\varrho}_{\mathrm{class},0}=\int d\boldsymbol{\rho}\exp\left(-\frac{2|\boldsymbol{\rho}|^2}{\sigma_-^2}\right)\left|\boldsymbol{\rho},\frac{\sigma_+}{\sqrt{2}},0\right\rangle\left|-\boldsymbol{\rho},\frac{\sigma_+}{\sqrt{2}},0\right\rangle\left\langle\boldsymbol{\rho},\frac{\sigma_+}{\sqrt{2}},0\right|\left\langle-\boldsymbol{\rho},\frac{\sigma_+}{\sqrt{2}},0\right|,
\end{equation}
where
\begin{equation}
    |\boldsymbol{\rho},\sigma,z\rangle=\int d\boldsymbol{\rho}'\exp\left(-\frac{|\boldsymbol{\rho}'-\boldsymbol{\rho}|^2}{\sigma^2+2iz/k}\right)|\boldsymbol{\rho}'\rangle
\end{equation}
is the Gaussian state with the center $\boldsymbol{\rho}$, the waist radius $\sigma$ and distance $z$. This is an ensemble of photon pairs having the same Gaussian spatial profile with opposite centers. We assume the photons are distinguishable (e.g. the two beams are spatially separated) and only consider $\sigma_+\ll\sigma_-$ here, so the photons are anticorrelated in position. The spatial JPD is
\begin{align}
    \langle\boldsymbol{\rho}_1|\langle\boldsymbol{\rho}_2|\hat{\varrho}_{\mathrm{class},0}|\boldsymbol{\rho}_1\rangle|\boldsymbol{\rho}_2\rangle&=\int d\boldsymbol{\rho}\exp\left(-\frac{2|\boldsymbol{\rho}|^2}{\sigma_-^2}-\frac{|\boldsymbol{\rho}_1-\boldsymbol{\rho}|^2+|\boldsymbol{\rho}_2+\boldsymbol{\rho}|^2}{\sigma_+^2}\right)\nonumber\\
    &\propto\exp\left[-\frac{2\sigma_+^2(|\boldsymbol{\rho}_1|^2+|\boldsymbol{\rho}_2|^2)+\sigma_-^2|\boldsymbol{\rho}_1+\boldsymbol{\rho}_2|^2}{2\sigma_+^2(\sigma_+^2+\sigma_-^2)}\right]\nonumber\\
    &\approx\exp\left(-\frac{|\boldsymbol{\rho}_1+\boldsymbol{\rho}_2|^2}{2\sigma_+^2}-\frac{|\boldsymbol{\rho}_1|^2+|\boldsymbol{\rho}_2|^2}{\sigma_-^2}\right)\nonumber\\
    &\approx\exp\left(-\frac{|\boldsymbol{\rho}_1+\boldsymbol{\rho}_2|^2}{2\sigma_+^2}-\frac{|\boldsymbol{\rho}_1-\boldsymbol{\rho}_2|^2}{2\sigma_-^2}\right),
\end{align}
which is the same as the entangled state represented by the squared modulus of Eq.\ (3) of the main text.

\begin{figure*}[h]
\includegraphics[width=0.7\textwidth]{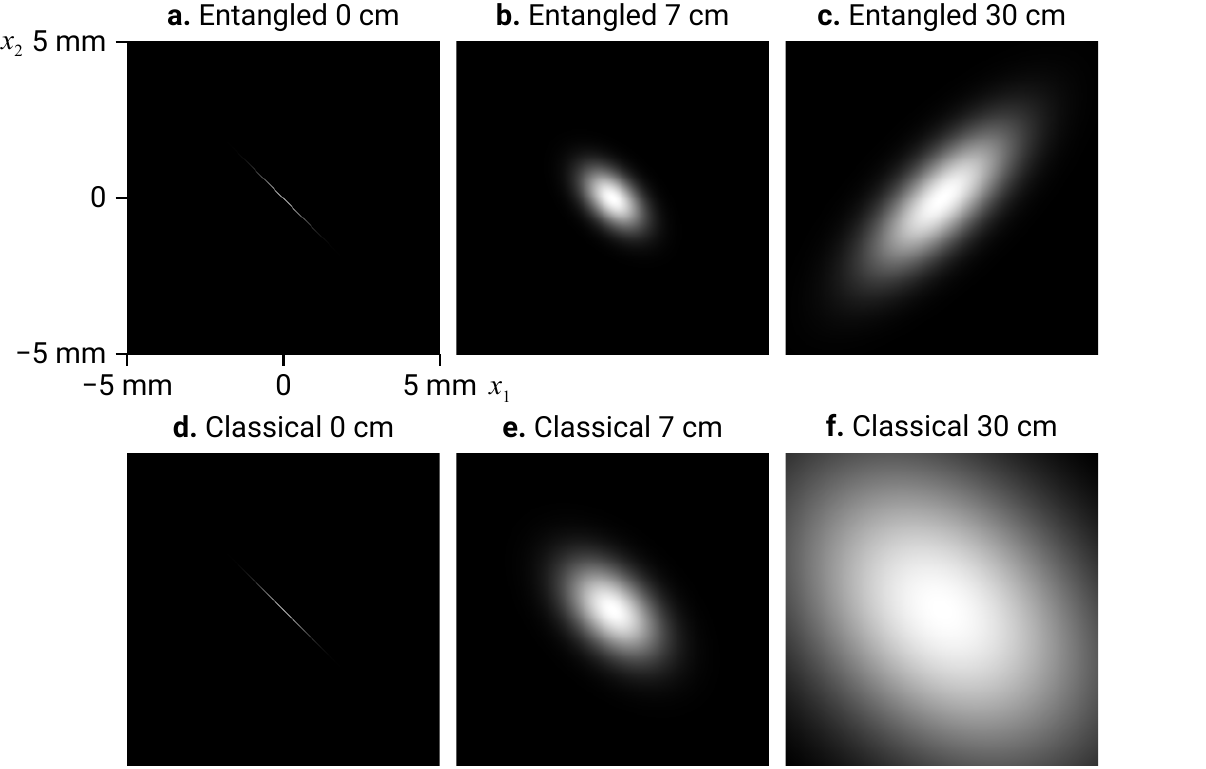}
\caption{\label{supclassfig}Comparison of the JPD of one-dimensional two-photon entangled and classically correlated state after propagating different distances in free space. The JPD is the same when there is no propagation ($z=0$).}
\end{figure*}

After propagating a distance $z$, the third parameters in the states in Eq.\ \eqref{classiform} are switched to $z$. In the one-dimensional case ($\boldsymbol{\rho}_1\to x_1,\boldsymbol{\rho}_2\to x_2$), We calculate the probability distribution first:
\begin{align}
    \langle x_1|\langle x_2|\hat{\varrho}_{\mathrm{class},z}|x_1\rangle|x_2\rangle&=\int dx\exp\left[-\frac{2x^2}{\sigma_-^2}-\frac{(x_1-x)^2+(x_2+x)^2}{\sigma_+^2+4z^2/(k\sigma_+)^2}\right]\nonumber\\
    &\propto\exp\left[-\frac{\sigma_-^2(x_1+x_2)^2}{2\left(\sigma_+^2+\frac{4z^2}{k^2\sigma_+^2}\right)\left(\sigma_+^2+\frac{4z^2}{k^2\sigma_+^2}+\sigma_-^2\right)}-\frac{x_1^2+x_2^2}{\sigma_+^2+\frac{4z^2}{k^2\sigma_+^2}+\sigma_-^2}\right].
\end{align}
After plotting (shown in Fig.\ \ref{supclassfig}), we find that as $z$ increases, the two photons are less anticorrelated, but they will never show positive correlation. When $z$ is very large, the photons at the far field has almost no position correlation.

\begin{figure*}[h]
\includegraphics[width=0.78\textwidth]{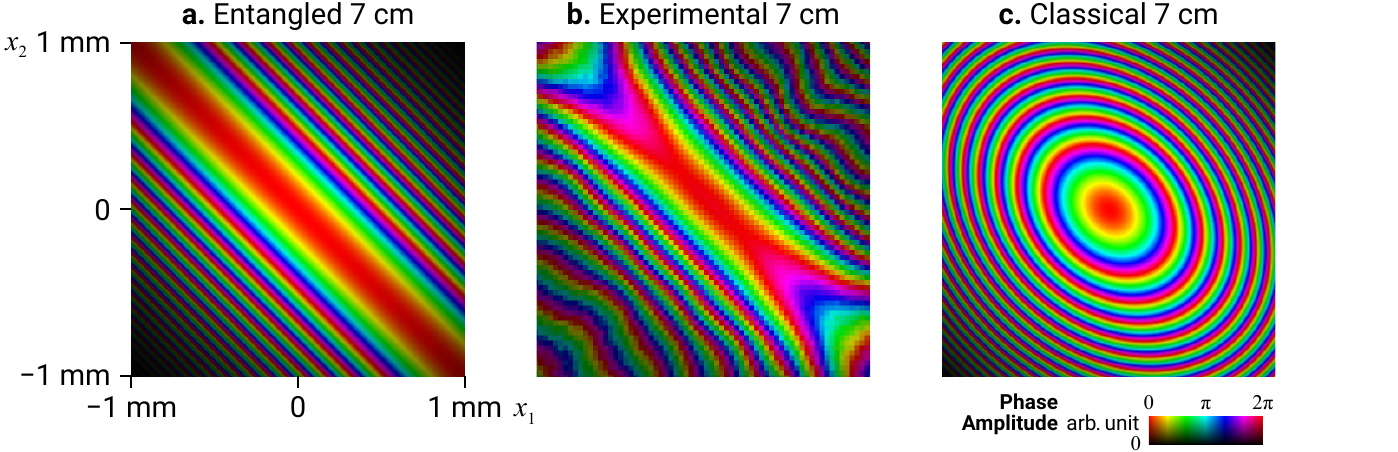}
\caption{\label{supcomparephase}Comparison of the joint wave function of entangled photon pair and the reconstructed ``wave function'' from the QSHWS using the classically correlated photon pair propagated by $7$~cm. A slice of the experimentally reconstructed wave function of $x_1,x_2$ ($y_1,y_2$ are near the center) is also given.}
\end{figure*}

Now we consider the QSHWS of this mixed state. Each pure component $\left|x,\sigma_+/\sqrt{2},z\right\rangle\left|-x,\sigma_+/\sqrt{2},z\right\rangle$ has an uncorrelated position JPD contribution on the back focal plane of the microlens array, and the JPDs of different components of the mixed state are directly summed. Each component has a different phase gradient distribution, so if we obtain the CPD of an accurate point at the sensor, one can tell which component it is from. However, if we calculate the CPD of an aperture, the spots in the CPD have sizes much larger than the usual Airy spot. If the average positions of the spots are used to calculate the so-called ``phase gradient'' (In the one-dimensional case, it is just the partial derivative), assuming the microlens is small, and letting $\varrho_{\mathrm{class},z}(x'_1,x'_2,x_1,x_2)=\langle x'_1|\langle x'_2|\hat{\varrho}_{\mathrm{class},z}|x_1\rangle|x_2\rangle$ (second-order coherence function), the measured ``phase gradient'' of photon 2 with a given position of photon 1 is [2,3]
\begin{align}
    \frac{\partial}{\partial x'_2}\operatorname{arg}\varrho_{\mathrm{class},z}(x_1,x'_2,x_1,x_2)\bigg|_{x'_2=x_2}&=8kz\left[\frac{x_1+x_2}{k^2\sigma_+^4+16z^2}-\frac{x_1-x_2}{k^2\sigma_+^2(\sigma_+^2+4\sigma_-^2)+16z^2}\right]\nonumber\\
    &\approx\frac{kz}{2}\left(\frac{x_1+x_2}{k^2\sigma_+^4/16+z^2}-\frac{x_1-x_2}{k^2\sigma_+^2\sigma_-^2/4+z^2}\right),
\end{align}
Given $x_1$, the ``gradient'' is still a linear function of $x_2$, so the ``phase distribution'' also has a quadratic form. Compared to the entangled case
\begin{equation}
    \frac{\partial}{\partial x_2}\operatorname{Im}\left[-\frac{(x_1+x_2)^2}{4(\sigma_+^2+iz/k)}-\frac{(x_1-x_2)^2}{4(\sigma_-^2+iz/k)}\right]=\frac{kz}{2}\left(\frac{x_1+x_2}{k^2\sigma_+^4+z^2}-\frac{x_1-x_2}{k^2\sigma_-^4+z^2}\right),
\end{equation}
when $\lambda=810~\mathrm{nm}$, $\sigma_\pm$ take the theoretical values in our experiment and $z=7~\mathrm{cm}$, the joint wave function of the entangled photons and the reconstructed ``wave function'' of the classically correlated photons are shown in Fig.\ \ref{supcomparephase}, where the experimentally reconstructed wave function using the $7$-cm data is also given. When the given point $x_1$ changes, the center of the parabolic ``phase'' pattern of the conditional ``wave function'' moves backward slower than $x_1$, while for the entangled case, the center of phase pattern moves almost at the same speed. These features make the classically correlated state different from the entangled state. The single-pixel CPD may provide more information about the properties of the unknown two-photon field, and this can be further explored in the future.

\newpage

\vspace{1em}
\centerline{\bf K.\;Derivation of the convolution formula}
\vspace{1em}

Here we provide a derivation of Eq.\ (5) in the main text:
\begin{align}  
    &\left[\exp\left(-\frac{|\boldsymbol{\rho}_1|^2}{\sigma_-^2}\right)\delta(\boldsymbol{\rho}_1+\boldsymbol{\rho}_2)\right]\ast\left[G(\boldsymbol{\rho}_1)G(\boldsymbol{\rho}_2)\right]\nonumber\\
    =&\int d\boldsymbol{\rho}'_1d\boldsymbol{\rho}'_2\exp\left(-\frac{|\boldsymbol{\rho}'_1|^2}{\sigma_-^2}\right)\delta(\boldsymbol{\rho}'_1+\boldsymbol{\rho}'_2)G(\boldsymbol{\rho}_1-\boldsymbol{\rho}'_1)G(\boldsymbol{\rho}_2-\boldsymbol{\rho}'_2)\nonumber\\
    =&\int d\boldsymbol{\rho}'\exp\left(-\frac{|\boldsymbol{\rho}'|^2}{\sigma_-^2}\right)G(\boldsymbol{\rho}_1-\boldsymbol{\rho}')G(\boldsymbol{\rho}_2+\boldsymbol{\rho}')\nonumber\\
    =&\int d\boldsymbol{\rho}'_2G(\boldsymbol{\rho}'_2)\exp\left(-\frac{|\boldsymbol{\rho}_1-\boldsymbol{\rho}'_2|^2}{\sigma_-^2}\right)G(\boldsymbol{\rho}_1+\boldsymbol{\rho}_2-\boldsymbol{\rho}'_2)\nonumber\\
    =&\left[G(\boldsymbol{\rho}_2)\exp\left(-\frac{|\boldsymbol{\rho}_1-\boldsymbol{\rho}_2|^2}{\sigma_-^2}\right)\right]\ast^{(\boldsymbol{\rho}_2)}G(\boldsymbol{\rho}_1+\boldsymbol{\rho}_2).
\end{align}
The convolution sign on the last line only acts on $\boldsymbol{\rho}_2$ ($\boldsymbol{\rho}_1$ is treated as a constant).

\vspace{2em}
\centerline{\bf References}
\vspace{1em}
\small

[1] Jorge Ares, Teresa Mancebo, and Salvador Bar\'{a}, Position and displacement sensing with Shack–Hartmann wave-front sensors, \href{https://doi.org/10.1364/AO.39.001511}{Appl.\ Opt.\ \textbf{39}, 1511 (2000)}.

[2] Y.\ Zheng, M.\ Yang, Z.-H.\ Liu, J.-S.\ Xu, C.-F.\ Li, and G.-C.\ Guo, Detecting momentum weak value: Shack–Hartmann versus a weak measurement wavefront sensor, \href{https://doi.org/10.1364/OL.439174}{Opt.\ Lett.\ \textbf{46}, 5352 (2021)}.

[3] Y.\ Zheng, M.\ Yang, Y.-W.\ Liao, J.-S.\ Xu, C.-F.\ Li, and G.-C.\ Guo, Reconstructing the multiphoton spatial wave function with coincidence wave-front sensing, \href{https://doi.org/10.1103/PhysRevA.107.042608}{Phys.\ Rev.\ A \textbf{107}, 042608 (2023)}.

[4] H.\ Defienne, M.\ Reichert, and J.\ W.\ Fleischer, General Model of Photon-Pair Detection with an Image Sensor, \href{https://doi.org/10.1103/PhysRevLett.120.203604}{Phys.\ Rev.\ Lett.\ \textbf{120}, 203604 (2018)}.

[5] M.\ Reichert, H.\ Defienne, and J.\ W.\ Fleischer, Optimizing the signal-to-noise ratio of biphoton distribution measurements, \href{https://doi.org/10.1103/PhysRevA.98.013841}{Phys.\ Rev.\ A \textbf{98}, 013841 (2018)}.

[6] H.\ Defienne, M.\ Reichert, J.\ W.\ Fleischer, and D.\ Faccio, Quantum image distillation, \href{https://doi.org/10.1126/sciadv.aax0307}{Sci.\ Adv.\ \textbf{5}, eaax0307 (2019)}.

[7] H.\ Defienne, B.\ Ndagano, A.\ Lyons, and D.\ Faccio, Polarization entanglement-enabled quantum holography, \href{https://doi.org/10.1038/s41567-020-01156-1}{Nat.\ Phys.\ \textbf{17}, 591 (2021)}.

[8] S.\ Walborn, C.\ Monken, S.\ P\'{a}dua, and P.\ Souto Ribeiro, Spatial correlations in parametric down-conversion, \href{https://doi.org/10.1016/j.physrep.2010.06.003}{Phys.\ Rep. \textbf{495}, 87 (2010)}.

[9] D.\ Eimerl, L.\ Davis, S.\ Velsko, E.\ K.\ Graham and A.\ Zalkin, Optical, mechanical, and thermal properties of barium borate, \href{https://doi.org/10.1063/1.339536}{J.\ Appl.\ Phys. \textbf{62}, 1968 (1987)}.

[10] K.\ W.\ Chan, J.\ P.\ Torres, and J.\ H.\ Eberly, Transverse entanglement migration in Hilbert space, \href{https://doi.org/10.1103/PhysRevA.75.050101}{Phys. Rev. A \textbf{75}, 050101 (2007)}.
\end{spacing}

\begin{thebibliography}{56}%
\makeatletter
\providecommand \@ifxundefined [1]{%
 \@ifx{#1\undefined}
}%
\providecommand \@ifnum [1]{%
 \ifnum #1\expandafter \@firstoftwo
 \else \expandafter \@secondoftwo
 \fi
}%
\providecommand \@ifx [1]{%
 \ifx #1\expandafter \@firstoftwo
 \else \expandafter \@secondoftwo
 \fi
}%
\providecommand \natexlab [1]{#1}%
\providecommand \enquote  [1]{``#1''}%
\providecommand \bibnamefont  [1]{#1}%
\providecommand \bibfnamefont [1]{#1}%
\providecommand \citenamefont [1]{#1}%
\providecommand \href@noop [0]{\@secondoftwo}%
\providecommand \href [0]{\begingroup \@sanitize@url \@href}%
\providecommand \@href[1]{\@@startlink{#1}\@@href}%
\providecommand \@@href[1]{\endgroup#1\@@endlink}%
\providecommand \@sanitize@url [0]{\catcode `\\12\catcode `\$12\catcode
  `\&12\catcode `\#12\catcode `\^12\catcode `\_12\catcode `\%12\relax}%
\providecommand \@@startlink[1]{}%
\providecommand \@@endlink[0]{}%
\providecommand \url  [0]{\begingroup\@sanitize@url \@url }%
\providecommand \@url [1]{\endgroup\@href {#1}{\urlprefix }}%
\providecommand \urlprefix  [0]{URL }%
\providecommand \Eprint [0]{\href }%
\providecommand \doibase [0]{https://doi.org/}%
\providecommand \selectlanguage [0]{\@gobble}%
\providecommand \bibinfo  [0]{\@secondoftwo}%
\providecommand \bibfield  [0]{\@secondoftwo}%
\providecommand \translation [1]{[#1]}%
\providecommand \BibitemOpen [0]{}%
\providecommand \bibitemStop [0]{}%
\providecommand \bibitemNoStop [0]{.\EOS\space}%
\providecommand \EOS [0]{\spacefactor3000\relax}%
\providecommand \BibitemShut  [1]{\csname bibitem#1\endcsname}%
\let\auto@bib@innerbib\@empty
\bibitem [{\citenamefont {Flamini}\ \emph {et~al.}(2018)\citenamefont
  {Flamini}, \citenamefont {Spagnolo},\ and\ \citenamefont
  {Sciarrino}}]{photonrev}%
  \BibitemOpen
  \bibfield  {author} {\bibinfo {author} {\bibfnamefont {F.}~\bibnamefont
  {Flamini}}, \bibinfo {author} {\bibfnamefont {N.}~\bibnamefont {Spagnolo}},\
  and\ \bibinfo {author} {\bibfnamefont {F.}~\bibnamefont {Sciarrino}},\
  }\bibfield  {title} {\bibinfo {title} {Photonic quantum information
  processing: a review},\ }\href {https://doi.org/10.1088/1361-6633/aad5b2}
  {\bibfield  {journal} {\bibinfo  {journal} {Rep. Prog. Phys.}\ }\textbf
  {\bibinfo {volume} {82}},\ \bibinfo {pages} {016001} (\bibinfo {year}
  {2018})}\BibitemShut {NoStop}%
\bibitem [{\citenamefont {Magaña-Loaiza}\ and\ \citenamefont
  {Boyd}(2019)}]{contrev}%
  \BibitemOpen
  \bibfield  {author} {\bibinfo {author} {\bibfnamefont {O.~S.}\ \bibnamefont
  {Magaña-Loaiza}}\ and\ \bibinfo {author} {\bibfnamefont {R.~W.}\
  \bibnamefont {Boyd}},\ }\bibfield  {title} {\bibinfo {title} {Quantum imaging
  and information},\ }\href {https://doi.org/10.1088/1361-6633/ab5005}
  {\bibfield  {journal} {\bibinfo  {journal} {Rep. Prog. Phys.}\ }\textbf
  {\bibinfo {volume} {82}},\ \bibinfo {pages} {124401} (\bibinfo {year}
  {2019})}\BibitemShut {NoStop}%
\bibitem [{\citenamefont {Pittman}\ \emph {et~al.}(1995)\citenamefont
  {Pittman}, \citenamefont {Shih}, \citenamefont {Strekalov},\ and\
  \citenamefont {Sergienko}}]{ghostimaging}%
  \BibitemOpen
  \bibfield  {author} {\bibinfo {author} {\bibfnamefont {T.~B.}\ \bibnamefont
  {Pittman}}, \bibinfo {author} {\bibfnamefont {Y.~H.}\ \bibnamefont {Shih}},
  \bibinfo {author} {\bibfnamefont {D.~V.}\ \bibnamefont {Strekalov}},\ and\
  \bibinfo {author} {\bibfnamefont {A.~V.}\ \bibnamefont {Sergienko}},\
  }\bibfield  {title} {\bibinfo {title} {Optical imaging by means of two-photon
  quantum entanglement},\ }\href {https://doi.org/10.1103/PhysRevA.52.R3429}
  {\bibfield  {journal} {\bibinfo  {journal} {Phys. Rev. A}\ }\textbf {\bibinfo
  {volume} {52}},\ \bibinfo {pages} {R3429} (\bibinfo {year}
  {1995})}\BibitemShut {NoStop}%
\bibitem [{\citenamefont {Lemos}\ \emph {et~al.}(2014)\citenamefont {Lemos},
  \citenamefont {Borish}, \citenamefont {Cole}, \citenamefont {Ramelow},
  \citenamefont {Lapkiewicz},\ and\ \citenamefont {Zeilinger}}]{Lemos2014}%
  \BibitemOpen
  \bibfield  {author} {\bibinfo {author} {\bibfnamefont {G.~B.}\ \bibnamefont
  {Lemos}}, \bibinfo {author} {\bibfnamefont {V.}~\bibnamefont {Borish}},
  \bibinfo {author} {\bibfnamefont {G.~D.}\ \bibnamefont {Cole}}, \bibinfo
  {author} {\bibfnamefont {S.}~\bibnamefont {Ramelow}}, \bibinfo {author}
  {\bibfnamefont {R.}~\bibnamefont {Lapkiewicz}},\ and\ \bibinfo {author}
  {\bibfnamefont {A.}~\bibnamefont {Zeilinger}},\ }\bibfield  {title} {\bibinfo
  {title} {Quantum imaging with undetected photons},\ }\href
  {https://doi.org/10.1038/nature13586} {\bibfield  {journal} {\bibinfo
  {journal} {Nature (London)}\ }\textbf {\bibinfo {volume} {512}},\ \bibinfo
  {pages} {409} (\bibinfo {year} {2014})}\BibitemShut {NoStop}%
\bibitem [{\citenamefont {Defienne}\ \emph {et~al.}(2019)\citenamefont
  {Defienne}, \citenamefont {Reichert}, \citenamefont {Fleischer},\ and\
  \citenamefont {Faccio}}]{distill}%
  \BibitemOpen
  \bibfield  {author} {\bibinfo {author} {\bibfnamefont {H.}~\bibnamefont
  {Defienne}}, \bibinfo {author} {\bibfnamefont {M.}~\bibnamefont {Reichert}},
  \bibinfo {author} {\bibfnamefont {J.~W.}\ \bibnamefont {Fleischer}},\ and\
  \bibinfo {author} {\bibfnamefont {D.}~\bibnamefont {Faccio}},\ }\bibfield
  {title} {\bibinfo {title} {Quantum image distillation},\ }\href
  {https://doi.org/10.1126/sciadv.aax0307} {\bibfield  {journal} {\bibinfo
  {journal} {Sci. Adv.}\ }\textbf {\bibinfo {volume} {5}},\ \bibinfo {pages}
  {eaax0307} (\bibinfo {year} {2019})}\BibitemShut {NoStop}%
\bibitem [{\citenamefont {Giovannetti}\ \emph {et~al.}(2009)\citenamefont
  {Giovannetti}, \citenamefont {Lloyd}, \citenamefont {Maccone},\ and\
  \citenamefont {Shapiro}}]{PhysRevA.79.013827}%
  \BibitemOpen
  \bibfield  {author} {\bibinfo {author} {\bibfnamefont {V.}~\bibnamefont
  {Giovannetti}}, \bibinfo {author} {\bibfnamefont {S.}~\bibnamefont {Lloyd}},
  \bibinfo {author} {\bibfnamefont {L.}~\bibnamefont {Maccone}},\ and\ \bibinfo
  {author} {\bibfnamefont {J.~H.}\ \bibnamefont {Shapiro}},\ }\bibfield
  {title} {\bibinfo {title} {{Sub-Rayleigh-diffraction-bound quantum
  imaging}},\ }\href {https://doi.org/10.1103/PhysRevA.79.013827} {\bibfield
  {journal} {\bibinfo  {journal} {Phys. Rev. A}\ }\textbf {\bibinfo {volume}
  {79}},\ \bibinfo {pages} {013827} (\bibinfo {year} {2009})}\BibitemShut
  {NoStop}%
\bibitem [{\citenamefont {He}\ \emph {et~al.}(2023)\citenamefont {He},
  \citenamefont {Zhang}, \citenamefont {Tong}, \citenamefont {Li},\ and\
  \citenamefont {Wang}}]{He2023}%
  \BibitemOpen
  \bibfield  {author} {\bibinfo {author} {\bibfnamefont {Z.}~\bibnamefont
  {He}}, \bibinfo {author} {\bibfnamefont {Y.}~\bibnamefont {Zhang}}, \bibinfo
  {author} {\bibfnamefont {X.}~\bibnamefont {Tong}}, \bibinfo {author}
  {\bibfnamefont {L.}~\bibnamefont {Li}},\ and\ \bibinfo {author}
  {\bibfnamefont {L.~V.}\ \bibnamefont {Wang}},\ }\bibfield  {title} {\bibinfo
  {title} {{Quantum microscopy of cells at the Heisenberg limit}},\ }\href
  {https://doi.org/10.1038/s41467-023-38191-4} {\bibfield  {journal} {\bibinfo
  {journal} {Nat. Commun.}\ }\textbf {\bibinfo {volume} {14}},\ \bibinfo
  {pages} {2441} (\bibinfo {year} {2023})}\BibitemShut {NoStop}%
\bibitem [{\citenamefont {Pan}\ \emph {et~al.}(2019)\citenamefont {Pan},
  \citenamefont {Xu}, \citenamefont {Kedem}, \citenamefont {Wang},
  \citenamefont {Chen}, \citenamefont {Jan}, \citenamefont {Sun}, \citenamefont
  {Xu}, \citenamefont {Han}, \citenamefont {Li},\ and\ \citenamefont
  {Guo}}]{PhysRevLett.123.150402}%
  \BibitemOpen
  \bibfield  {author} {\bibinfo {author} {\bibfnamefont {W.-W.}\ \bibnamefont
  {Pan}}, \bibinfo {author} {\bibfnamefont {X.-Y.}\ \bibnamefont {Xu}},
  \bibinfo {author} {\bibfnamefont {Y.}~\bibnamefont {Kedem}}, \bibinfo
  {author} {\bibfnamefont {Q.-Q.}\ \bibnamefont {Wang}}, \bibinfo {author}
  {\bibfnamefont {Z.}~\bibnamefont {Chen}}, \bibinfo {author} {\bibfnamefont
  {M.}~\bibnamefont {Jan}}, \bibinfo {author} {\bibfnamefont {K.}~\bibnamefont
  {Sun}}, \bibinfo {author} {\bibfnamefont {J.-S.}\ \bibnamefont {Xu}},
  \bibinfo {author} {\bibfnamefont {Y.-J.}\ \bibnamefont {Han}}, \bibinfo
  {author} {\bibfnamefont {C.-F.}\ \bibnamefont {Li}},\ and\ \bibinfo {author}
  {\bibfnamefont {G.-C.}\ \bibnamefont {Guo}},\ }\bibfield  {title} {\bibinfo
  {title} {{Direct Measurement of a Nonlocal Entangled Quantum State}},\ }\href
  {https://doi.org/10.1103/PhysRevLett.123.150402} {\bibfield  {journal}
  {\bibinfo  {journal} {Phys. Rev. Lett.}\ }\textbf {\bibinfo {volume} {123}},\
  \bibinfo {pages} {150402} (\bibinfo {year} {2019})}\BibitemShut {NoStop}%
\bibitem [{\citenamefont {Lundeen}\ \emph {et~al.}(2011)\citenamefont
  {Lundeen}, \citenamefont {Sutherland}, \citenamefont {Patel}, \citenamefont
  {Stewart},\ and\ \citenamefont {Bamber}}]{Lundeen2011}%
  \BibitemOpen
  \bibfield  {author} {\bibinfo {author} {\bibfnamefont {J.~S.}\ \bibnamefont
  {Lundeen}}, \bibinfo {author} {\bibfnamefont {B.}~\bibnamefont {Sutherland}},
  \bibinfo {author} {\bibfnamefont {A.}~\bibnamefont {Patel}}, \bibinfo
  {author} {\bibfnamefont {C.}~\bibnamefont {Stewart}},\ and\ \bibinfo {author}
  {\bibfnamefont {C.}~\bibnamefont {Bamber}},\ }\bibfield  {title} {\bibinfo
  {title} {Direct measurement of the quantum wavefunction},\ }\href
  {https://doi.org/10.1038/nature10120} {\bibfield  {journal} {\bibinfo
  {journal} {Nature (London)}\ }\textbf {\bibinfo {volume} {474}},\ \bibinfo
  {pages} {188} (\bibinfo {year} {2011})}\BibitemShut {NoStop}%
\bibitem [{\citenamefont {Howell}\ \emph {et~al.}(2004)\citenamefont {Howell},
  \citenamefont {Bennink}, \citenamefont {Bentley},\ and\ \citenamefont
  {Boyd}}]{PhysRevLett.92.210403}%
  \BibitemOpen
  \bibfield  {author} {\bibinfo {author} {\bibfnamefont {J.~C.}\ \bibnamefont
  {Howell}}, \bibinfo {author} {\bibfnamefont {R.~S.}\ \bibnamefont {Bennink}},
  \bibinfo {author} {\bibfnamefont {S.~J.}\ \bibnamefont {Bentley}},\ and\
  \bibinfo {author} {\bibfnamefont {R.~W.}\ \bibnamefont {Boyd}},\ }\bibfield
  {title} {\bibinfo {title} {{Realization of the Einstein-Podolsky-Rosen
  Paradox Using Momentum- and Position-Entangled Photons from Spontaneous
  Parametric Down Conversion}},\ }\href
  {https://doi.org/10.1103/PhysRevLett.92.210403} {\bibfield  {journal}
  {\bibinfo  {journal} {Phys. Rev. Lett.}\ }\textbf {\bibinfo {volume} {92}},\
  \bibinfo {pages} {210403} (\bibinfo {year} {2004})}\BibitemShut {NoStop}%
\bibitem [{\citenamefont {Wang}\ \emph {et~al.}(2012)\citenamefont {Wang},
  \citenamefont {Yang}, \citenamefont {Fazal}, \citenamefont {Ahmed},
  \citenamefont {Yan}, \citenamefont {Huang}, \citenamefont {Ren},
  \citenamefont {Yue}, \citenamefont {Dolinar}, \citenamefont {Tur},\ and\
  \citenamefont {Willner}}]{Wang2012}%
  \BibitemOpen
  \bibfield  {author} {\bibinfo {author} {\bibfnamefont {J.}~\bibnamefont
  {Wang}}, \bibinfo {author} {\bibfnamefont {J.-Y.}\ \bibnamefont {Yang}},
  \bibinfo {author} {\bibfnamefont {I.~M.}\ \bibnamefont {Fazal}}, \bibinfo
  {author} {\bibfnamefont {N.}~\bibnamefont {Ahmed}}, \bibinfo {author}
  {\bibfnamefont {Y.}~\bibnamefont {Yan}}, \bibinfo {author} {\bibfnamefont
  {H.}~\bibnamefont {Huang}}, \bibinfo {author} {\bibfnamefont
  {Y.}~\bibnamefont {Ren}}, \bibinfo {author} {\bibfnamefont {Y.}~\bibnamefont
  {Yue}}, \bibinfo {author} {\bibfnamefont {S.}~\bibnamefont {Dolinar}},
  \bibinfo {author} {\bibfnamefont {M.}~\bibnamefont {Tur}},\ and\ \bibinfo
  {author} {\bibfnamefont {A.~E.}\ \bibnamefont {Willner}},\ }\bibfield
  {title} {\bibinfo {title} {Terabit free-space data transmission employing
  orbital angular momentum multiplexing},\ }\href
  {https://doi.org/10.1038/nphoton.2012.138} {\bibfield  {journal} {\bibinfo
  {journal} {Nat. Photonics}\ }\textbf {\bibinfo {volume} {6}},\ \bibinfo {pages}
  {488} (\bibinfo {year} {2012})}\BibitemShut {NoStop}%
\bibitem [{\citenamefont {Park}\ \emph {et~al.}(2018)\citenamefont {Park},
  \citenamefont {Depeursinge},\ and\ \citenamefont {Popescu}}]{Park2018}%
  \BibitemOpen
  \bibfield  {author} {\bibinfo {author} {\bibfnamefont {Y.}~\bibnamefont
  {Park}}, \bibinfo {author} {\bibfnamefont {C.}~\bibnamefont {Depeursinge}},\
  and\ \bibinfo {author} {\bibfnamefont {G.}~\bibnamefont {Popescu}},\
  }\bibfield  {title} {\bibinfo {title} {Quantitative phase imaging in
  biomedicine},\ }\href {https://doi.org/10.1038/s41566-018-0253-x} {\bibfield
  {journal} {\bibinfo  {journal} {Nat. Photonics}\ }\textbf {\bibinfo {volume}
  {12}},\ \bibinfo {pages} {578} (\bibinfo {year} {2018})}\BibitemShut
  {NoStop}%
\bibitem [{\citenamefont {Gabor}(1948)}]{GABOR1948}%
  \BibitemOpen
  \bibfield  {author} {\bibinfo {author} {\bibfnamefont {D.}~\bibnamefont
  {Gabor}},\ }\bibfield  {title} {\bibinfo {title} {{A New Microscopic
  Principle}},\ }\href {https://doi.org/10.1038/161777a0} {\bibfield  {journal}
  {\bibinfo  {journal} {Nature (London)}\ }\textbf {\bibinfo {volume} {161}},\
  \bibinfo {pages} {777} (\bibinfo {year} {1948})}\BibitemShut {NoStop}%
\bibitem [{\citenamefont {Yamaguchi}\ and\ \citenamefont
  {Zhang}(1997)}]{Yamaguchi:97}%
  \BibitemOpen
  \bibfield  {author} {\bibinfo {author} {\bibfnamefont {I.}~\bibnamefont
  {Yamaguchi}}\ and\ \bibinfo {author} {\bibfnamefont {T.}~\bibnamefont
  {Zhang}},\ }\bibfield  {title} {\bibinfo {title} {Phase-shifting digital
  holography},\ }\href {https://doi.org/10.1364/OL.22.001268} {\bibfield
  {journal} {\bibinfo  {journal} {Opt. Lett.}\ }\textbf {\bibinfo {volume}
  {22}},\ \bibinfo {pages} {1268} (\bibinfo {year} {1997})}\BibitemShut
  {NoStop}%
\bibitem [{\citenamefont {Defienne}\ \emph {et~al.}(2021)\citenamefont
  {Defienne}, \citenamefont {Ndagano}, \citenamefont {Lyons},\ and\
  \citenamefont {Faccio}}]{Defienne2021}%
  \BibitemOpen
  \bibfield  {author} {\bibinfo {author} {\bibfnamefont {H.}~\bibnamefont
  {Defienne}}, \bibinfo {author} {\bibfnamefont {B.}~\bibnamefont {Ndagano}},
  \bibinfo {author} {\bibfnamefont {A.}~\bibnamefont {Lyons}},\ and\ \bibinfo
  {author} {\bibfnamefont {D.}~\bibnamefont {Faccio}},\ }\bibfield  {title}
  {\bibinfo {title} {Polarization entanglement-enabled quantum holography},\
  }\href {https://doi.org/10.1038/s41567-020-01156-1} {\bibfield  {journal}
  {\bibinfo  {journal} {Nat. Phys.}\ }\textbf {\bibinfo {volume} {17}},\
  \bibinfo {pages} {591} (\bibinfo {year} {2021})}\BibitemShut {NoStop}%
\bibitem [{\citenamefont {Camphausen}\ \emph {et~al.}(2021)\citenamefont
  {Camphausen}, \citenamefont {Álvaro Cuevas}, \citenamefont {Duempelmann},
  \citenamefont {Terborg}, \citenamefont {Wajs}, \citenamefont {Tisa},
  \citenamefont {Ruggeri}, \citenamefont {Cusini}, \citenamefont
  {Steinlechner},\ and\ \citenamefont {Pruneri}}]{sciadv.abj2155}%
  \BibitemOpen
  \bibfield  {author} {\bibinfo {author} {\bibfnamefont {R.}~\bibnamefont
  {Camphausen}}, \bibinfo {author} {\bibnamefont {Álvaro Cuevas}}, \bibinfo
  {author} {\bibfnamefont {L.}~\bibnamefont {Duempelmann}}, \bibinfo {author}
  {\bibfnamefont {R.~A.}\ \bibnamefont {Terborg}}, \bibinfo {author}
  {\bibfnamefont {E.}~\bibnamefont {Wajs}}, \bibinfo {author} {\bibfnamefont
  {S.}~\bibnamefont {Tisa}}, \bibinfo {author} {\bibfnamefont {A.}~\bibnamefont
  {Ruggeri}}, \bibinfo {author} {\bibfnamefont {I.}~\bibnamefont {Cusini}},
  \bibinfo {author} {\bibfnamefont {F.}~\bibnamefont {Steinlechner}},\ and\
  \bibinfo {author} {\bibfnamefont {V.}~\bibnamefont {Pruneri}},\ }\bibfield
  {title} {\bibinfo {title} {A quantum-enhanced wide-field phase imager},\
  }\href {https://doi.org/10.1126/sciadv.abj2155} {\bibfield  {journal}
  {\bibinfo  {journal} {Sci. Adv.}\ }\textbf {\bibinfo {volume} {7}},\ \bibinfo
  {pages} {eabj2155} (\bibinfo {year} {2021})}\BibitemShut {NoStop}%
\bibitem [{\citenamefont {Zia}\ \emph {et~al.}(2023)\citenamefont {Zia},
  \citenamefont {Dehghan}, \citenamefont {D'Errico}, \citenamefont
  {Sciarrino},\ and\ \citenamefont {Karimi}}]{Zia2023}%
  \BibitemOpen
  \bibfield  {author} {\bibinfo {author} {\bibfnamefont {D.}~\bibnamefont
  {Zia}}, \bibinfo {author} {\bibfnamefont {N.}~\bibnamefont {Dehghan}},
  \bibinfo {author} {\bibfnamefont {A.}~\bibnamefont {D'Errico}}, \bibinfo
  {author} {\bibfnamefont {F.}~\bibnamefont {Sciarrino}},\ and\ \bibinfo
  {author} {\bibfnamefont {E.}~\bibnamefont {Karimi}},\ }\bibfield  {title}
  {\bibinfo {title} {Interferometric imaging of amplitude and phase of spatial
  biphoton states},\ }\href {https://doi.org/10.1038/s41566-023-01272-3}
  {\bibfield  {journal} {\bibinfo  {journal} {Nat. Photonics}\ }\textbf {\bibinfo
  {volume} {17}},\ \bibinfo {pages} {1009} (\bibinfo {year}
  {2023})}\BibitemShut {NoStop}%
\bibitem [{\citenamefont {Zernike}(1955)}]{Zernikephase}%
  \BibitemOpen
  \bibfield  {author} {\bibinfo {author} {\bibfnamefont {F.}~\bibnamefont
  {Zernike}},\ }\bibfield  {title} {\bibinfo {title} {{How I Discovered Phase
  Contrast}},\ }\href {https://doi.org/10.1126/science.121.3141.345} {\bibfield
   {journal} {\bibinfo  {journal} {Science}\ }\textbf {\bibinfo {volume}
  {121}},\ \bibinfo {pages} {345} (\bibinfo {year} {1955})}\BibitemShut
  {NoStop}%
\bibitem [{\citenamefont {Dressel}\ \emph {et~al.}(2014)\citenamefont
  {Dressel}, \citenamefont {Malik}, \citenamefont {Miatto}, \citenamefont
  {Jordan},\ and\ \citenamefont {Boyd}}]{RevModPhys.86.307}%
  \BibitemOpen
  \bibfield  {author} {\bibinfo {author} {\bibfnamefont {J.}~\bibnamefont
  {Dressel}}, \bibinfo {author} {\bibfnamefont {M.}~\bibnamefont {Malik}},
  \bibinfo {author} {\bibfnamefont {F.~M.}\ \bibnamefont {Miatto}}, \bibinfo
  {author} {\bibfnamefont {A.~N.}\ \bibnamefont {Jordan}},\ and\ \bibinfo
  {author} {\bibfnamefont {R.~W.}\ \bibnamefont {Boyd}},\ }\bibfield  {title}
  {\bibinfo {title} {{Colloquium: Understanding quantum weak values: Basics and
  applications}},\ }\href {https://doi.org/10.1103/RevModPhys.86.307}
  {\bibfield  {journal} {\bibinfo  {journal} {Rev. Mod. Phys.}\ }\textbf
  {\bibinfo {volume} {86}},\ \bibinfo {pages} {307} (\bibinfo {year}
  {2014})}\BibitemShut {NoStop}%
\bibitem [{\citenamefont {Shi}\ \emph {et~al.}(2015)\citenamefont {Shi},
  \citenamefont {Mirhosseini}, \citenamefont {Margiewicz}, \citenamefont
  {Malik}, \citenamefont {Rivera}, \citenamefont {Zhu},\ and\ \citenamefont
  {Boyd}}]{Shi:15}%
  \BibitemOpen
  \bibfield  {author} {\bibinfo {author} {\bibfnamefont {Z.}~\bibnamefont
  {Shi}}, \bibinfo {author} {\bibfnamefont {M.}~\bibnamefont {Mirhosseini}},
  \bibinfo {author} {\bibfnamefont {J.}~\bibnamefont {Margiewicz}}, \bibinfo
  {author} {\bibfnamefont {M.}~\bibnamefont {Malik}}, \bibinfo {author}
  {\bibfnamefont {F.}~\bibnamefont {Rivera}}, \bibinfo {author} {\bibfnamefont
  {Z.}~\bibnamefont {Zhu}},\ and\ \bibinfo {author} {\bibfnamefont {R.~W.}\
  \bibnamefont {Boyd}},\ }\bibfield  {title} {\bibinfo {title} {Scan-free
  direct measurement of an extremely high-dimensional photonic state},\ }\href
  {https://doi.org/10.1364/OPTICA.2.000388} {\bibfield  {journal} {\bibinfo
  {journal} {Optica}\ }\textbf {\bibinfo {volume} {2}},\ \bibinfo {pages} {388}
  (\bibinfo {year} {2015})}\BibitemShut {NoStop}%
\bibitem [{\citenamefont {Kocsis}\ \emph {et~al.}(2011)\citenamefont {Kocsis},
  \citenamefont {Braverman}, \citenamefont {Ravets}, \citenamefont {Stevens},
  \citenamefont {Mirin}, \citenamefont {Shalm},\ and\ \citenamefont
  {Steinberg}}]{Kocsis1170}%
  \BibitemOpen
  \bibfield  {author} {\bibinfo {author} {\bibfnamefont {S.}~\bibnamefont
  {Kocsis}}, \bibinfo {author} {\bibfnamefont {B.}~\bibnamefont {Braverman}},
  \bibinfo {author} {\bibfnamefont {S.}~\bibnamefont {Ravets}}, \bibinfo
  {author} {\bibfnamefont {M.~J.}\ \bibnamefont {Stevens}}, \bibinfo {author}
  {\bibfnamefont {R.~P.}\ \bibnamefont {Mirin}}, \bibinfo {author}
  {\bibfnamefont {L.~K.}\ \bibnamefont {Shalm}},\ and\ \bibinfo {author}
  {\bibfnamefont {A.~M.}\ \bibnamefont {Steinberg}},\ }\bibfield  {title}
  {\bibinfo {title} {{Observing the Average Trajectories of Single Photons in a
  Two-Slit Interferometer}},\ }\href {https://doi.org/10.1126/science.1202218}
  {\bibfield  {journal} {\bibinfo  {journal} {Science}\ }\textbf {\bibinfo
  {volume} {332}},\ \bibinfo {pages} {1170} (\bibinfo {year}
  {2011})}\BibitemShut {NoStop}%
\bibitem [{\citenamefont {Yang}\ \emph {et~al.}(2020)\citenamefont {Yang},
  \citenamefont {Xiao}, \citenamefont {Liao}, \citenamefont {Liu},
  \citenamefont {Xu}, \citenamefont {Xu}, \citenamefont {Li},\ and\
  \citenamefont {Guo}}]{Yang2020}%
  \BibitemOpen
  \bibfield  {author} {\bibinfo {author} {\bibfnamefont {M.}~\bibnamefont
  {Yang}}, \bibinfo {author} {\bibfnamefont {Y.}~\bibnamefont {Xiao}}, \bibinfo
  {author} {\bibfnamefont {Y.-W.}\ \bibnamefont {Liao}}, \bibinfo {author}
  {\bibfnamefont {Z.-H.}\ \bibnamefont {Liu}}, \bibinfo {author} {\bibfnamefont
  {X.-Y.}\ \bibnamefont {Xu}}, \bibinfo {author} {\bibfnamefont {J.-S.}\
  \bibnamefont {Xu}}, \bibinfo {author} {\bibfnamefont {C.-F.}\ \bibnamefont
  {Li}},\ and\ \bibinfo {author} {\bibfnamefont {G.-C.}\ \bibnamefont {Guo}},\
  }\bibfield  {title} {\bibinfo {title} {{Zonal Reconstruction of Photonic
  Wavefunction via Momentum Weak Measurement}},\ }\href
  {https://doi.org/10.1002/lpor.201900251} {\bibfield  {journal} {\bibinfo
  {journal} {Laser Photonics Rev.}\ }\textbf {\bibinfo {volume} {14}},\ \bibinfo
  {pages} {1900251} (\bibinfo {year} {2020})}\BibitemShut {NoStop}%
\bibitem [{\citenamefont {Hudgin}(1977)}]{Hudgin:77}%
  \BibitemOpen
  \bibfield  {author} {\bibinfo {author} {\bibfnamefont {R.~H.}\ \bibnamefont
  {Hudgin}},\ }\bibfield  {title} {\bibinfo {title} {Wave-front reconstruction
  for compensated imaging},\ }\href {https://doi.org/10.1364/JOSA.67.000375}
  {\bibfield  {journal} {\bibinfo  {journal} {J. Opt. Soc. Am.}\ }\textbf
  {\bibinfo {volume} {67}},\ \bibinfo {pages} {375} (\bibinfo {year}
  {1977})}\BibitemShut {NoStop}%
\bibitem [{\citenamefont {Southwell}(1980)}]{Southwell:80}%
  \BibitemOpen
  \bibfield  {author} {\bibinfo {author} {\bibfnamefont {W.}~\bibnamefont
  {Southwell}},\ }\bibfield  {title} {\bibinfo {title} {Wave-front estimation
  from wave-front slope measurements},\ }\href
  {https://doi.org/10.1364/JOSA.70.000998} {\bibfield  {journal} {\bibinfo
  {journal} {J. Opt. Soc. Am.}\ }\textbf {\bibinfo {volume} {70}},\ \bibinfo
  {pages} {998} (\bibinfo {year} {1980})}\BibitemShut {NoStop}%
\bibitem [{\citenamefont {Zheng}\ \emph {et~al.}(2021)\citenamefont {Zheng},
  \citenamefont {Yang}, \citenamefont {Liu}, \citenamefont {Xu}, \citenamefont
  {Li},\ and\ \citenamefont {Guo}}]{Zheng:21}%
  \BibitemOpen
  \bibfield  {author} {\bibinfo {author} {\bibfnamefont {Y.}~\bibnamefont
  {Zheng}}, \bibinfo {author} {\bibfnamefont {M.}~\bibnamefont {Yang}},
  \bibinfo {author} {\bibfnamefont {Z.-H.}\ \bibnamefont {Liu}}, \bibinfo
  {author} {\bibfnamefont {J.-S.}\ \bibnamefont {Xu}}, \bibinfo {author}
  {\bibfnamefont {C.-F.}\ \bibnamefont {Li}},\ and\ \bibinfo {author}
  {\bibfnamefont {G.-C.}\ \bibnamefont {Guo}},\ }\bibfield  {title} {\bibinfo
  {title} {{Detecting momentum weak value: Shack--Hartmann versus a weak
  measurement wavefront sensor}},\ }\href {https://doi.org/10.1364/OL.439174}
  {\bibfield  {journal} {\bibinfo  {journal} {Opt. Lett.}\ }\textbf {\bibinfo
  {volume} {46}},\ \bibinfo {pages} {5352} (\bibinfo {year}
  {2021})}\BibitemShut {NoStop}%
\bibitem [{\citenamefont {Zheng}\ \emph {et~al.}(2022)\citenamefont {Zheng},
  \citenamefont {Yang}, \citenamefont {Liu}, \citenamefont {Xu}, \citenamefont
  {Li},\ and\ \citenamefont {Guo}}]{Zheng:22}%
  \BibitemOpen
  \bibfield  {author} {\bibinfo {author} {\bibfnamefont {Y.}~\bibnamefont
  {Zheng}}, \bibinfo {author} {\bibfnamefont {M.}~\bibnamefont {Yang}},
  \bibinfo {author} {\bibfnamefont {Z.-H.}\ \bibnamefont {Liu}}, \bibinfo
  {author} {\bibfnamefont {J.-S.}\ \bibnamefont {Xu}}, \bibinfo {author}
  {\bibfnamefont {C.-F.}\ \bibnamefont {Li}},\ and\ \bibinfo {author}
  {\bibfnamefont {G.-C.}\ \bibnamefont {Guo}},\ }\bibfield  {title} {\bibinfo
  {title} {Toward practical weak measurement wavefront sensing: spatial
  resolution and achromatism},\ }\href {https://doi.org/10.1364/OL.460873}
  {\bibfield  {journal} {\bibinfo  {journal} {Opt. Lett.}\ }\textbf {\bibinfo
  {volume} {47}},\ \bibinfo {pages} {2734} (\bibinfo {year}
  {2022})}\BibitemShut {NoStop}%
\bibitem [{\citenamefont {Zheng}\ \emph {et~al.}(2023)\citenamefont {Zheng},
  \citenamefont {Yang}, \citenamefont {Liao}, \citenamefont {Xu}, \citenamefont
  {Li},\ and\ \citenamefont {Guo}}]{Zheng2023}%
  \BibitemOpen
  \bibfield  {author} {\bibinfo {author} {\bibfnamefont {Y.}~\bibnamefont
  {Zheng}}, \bibinfo {author} {\bibfnamefont {M.}~\bibnamefont {Yang}},
  \bibinfo {author} {\bibfnamefont {Y.-W.}\ \bibnamefont {Liao}}, \bibinfo
  {author} {\bibfnamefont {J.-S.}\ \bibnamefont {Xu}}, \bibinfo {author}
  {\bibfnamefont {C.-F.}\ \bibnamefont {Li}},\ and\ \bibinfo {author}
  {\bibfnamefont {G.-C.}\ \bibnamefont {Guo}},\ }\bibfield  {title} {\bibinfo
  {title} {Reconstructing the multiphoton spatial wave function with
  coincidence wave-front sensing},\ }\href
  {https://doi.org/10.1103/PhysRevA.107.042608} {\bibfield  {journal} {\bibinfo
   {journal} {Phys. Rev. A}\ }\textbf {\bibinfo {volume} {107}},\ \bibinfo
  {pages} {042608} (\bibinfo {year} {2023})}\BibitemShut {NoStop}%
\bibitem [{\citenamefont {Hu}\ \emph {et~al.}(2020)\citenamefont {Hu},
  \citenamefont {Xing}, \citenamefont {Liu}, \citenamefont {Huang},
  \citenamefont {Li}, \citenamefont {Guo}, \citenamefont {Erker},\ and\
  \citenamefont {Huber}}]{PhysRevLett.125.090503}%
  \BibitemOpen
  \bibfield  {author} {\bibinfo {author} {\bibfnamefont {X.-M.}\ \bibnamefont
  {Hu}}, \bibinfo {author} {\bibfnamefont {W.-B.}\ \bibnamefont {Xing}},
  \bibinfo {author} {\bibfnamefont {B.-H.}\ \bibnamefont {Liu}}, \bibinfo
  {author} {\bibfnamefont {Y.-F.}\ \bibnamefont {Huang}}, \bibinfo {author}
  {\bibfnamefont {C.-F.}\ \bibnamefont {Li}}, \bibinfo {author} {\bibfnamefont
  {G.-C.}\ \bibnamefont {Guo}}, \bibinfo {author} {\bibfnamefont
  {P.}~\bibnamefont {Erker}},\ and\ \bibinfo {author} {\bibfnamefont
  {M.}~\bibnamefont {Huber}},\ }\bibfield  {title} {\bibinfo {title}
  {{Efficient Generation of High-Dimensional Entanglement through Multipath
  Down-Conversion}},\ }\href {https://doi.org/10.1103/PhysRevLett.125.090503}
  {\bibfield  {journal} {\bibinfo  {journal} {Phys. Rev. Lett.}\ }\textbf
  {\bibinfo {volume} {125}},\ \bibinfo {pages} {090503} (\bibinfo {year}
  {2020})}\BibitemShut {NoStop}%
\bibitem [{\citenamefont {Platt}\ and\ \citenamefont {Shack}(2001)}]{Shack01}%
  \BibitemOpen
  \bibfield  {author} {\bibinfo {author} {\bibfnamefont {B.~C.}\ \bibnamefont
  {Platt}}\ and\ \bibinfo {author} {\bibfnamefont {R.}~\bibnamefont {Shack}},\
  }\bibfield  {title} {\bibinfo {title} {{History and Principles of
  Shack-Hartmann Wavefront Sensing}},\ }\href
  {https://doi.org/10.3928/1081-597X-20010901-13} {\bibfield  {journal}
  {\bibinfo  {journal} {J. Refract. Surg.}\ }\textbf {\bibinfo {volume} {17}},\
  \bibinfo {pages} {S573} (\bibinfo {year} {2001})}\BibitemShut {NoStop}%
\bibitem [{\citenamefont {Ares}\ \emph {et~al.}(2000)\citenamefont {Ares},
  \citenamefont {Mancebo},\ and\ \citenamefont {Bar\'{a}}}]{Ares:00}%
  \BibitemOpen
  \bibfield  {author} {\bibinfo {author} {\bibfnamefont {J.}~\bibnamefont
  {Ares}}, \bibinfo {author} {\bibfnamefont {T.}~\bibnamefont {Mancebo}},\ and\
  \bibinfo {author} {\bibfnamefont {S.}~\bibnamefont {Bar\'{a}}},\ }\bibfield
  {title} {\bibinfo {title} {{Position and displacement sensing with
  Shack--Hartmann wave-front sensors}},\ }\href
  {https://doi.org/10.1364/AO.39.001511} {\bibfield  {journal} {\bibinfo
  {journal} {Appl. Opt.}\ }\textbf {\bibinfo {volume} {39}},\ \bibinfo {pages}
  {1511} (\bibinfo {year} {2000})}\BibitemShut {NoStop}%
\bibitem [{sm()}]{sm}%
  \BibitemOpen
  \href@noop {} {\bibinfo {title} {{See \hyperlink{smpage}{Supplemental Material} for details of the experimental setup, data
  processing, more figures, and theoretical analyses.}}}\BibitemShut {Stop}%
\bibitem [{\citenamefont {Walborn}\ \emph {et~al.}(2010)\citenamefont
  {Walborn}, \citenamefont {Monken}, \citenamefont {Pádua},\ and\
  \citenamefont {{Souto Ribeiro}}}]{WALBORN201087}%
  \BibitemOpen
  \bibfield  {author} {\bibinfo {author} {\bibfnamefont {S.}~\bibnamefont
  {Walborn}}, \bibinfo {author} {\bibfnamefont {C.}~\bibnamefont {Monken}},
  \bibinfo {author} {\bibfnamefont {S.}~\bibnamefont {Pádua}},\ and\ \bibinfo
  {author} {\bibfnamefont {P.}~\bibnamefont {{Souto Ribeiro}}},\ }\bibfield
  {title} {\bibinfo {title} {Spatial correlations in parametric
  down-conversion},\ }\href {https://doi.org/10.1016/j.physrep.2010.06.003}
  {\bibfield  {journal} {\bibinfo  {journal} {Phys. Rep.}\ }\textbf {\bibinfo
  {volume} {495}},\ \bibinfo {pages} {87} (\bibinfo {year} {2010})}\BibitemShut
  {NoStop}%
\bibitem [{\citenamefont {Schneeloch}\ and\ \citenamefont
  {Howell}(2016)}]{Schneeloch_2016}%
  \BibitemOpen
  \bibfield  {author} {\bibinfo {author} {\bibfnamefont {J.}~\bibnamefont
  {Schneeloch}}\ and\ \bibinfo {author} {\bibfnamefont {J.~C.}\ \bibnamefont
  {Howell}},\ }\bibfield  {title} {\bibinfo {title} {Introduction to the
  transverse spatial correlations in spontaneous parametric down-conversion
  through the biphoton birth zone},\ }\href
  {https://doi.org/10.1088/2040-8978/18/5/053501} {\bibfield  {journal}
  {\bibinfo  {journal} {J. Opt.}\ }\textbf {\bibinfo {volume} {18}},\ \bibinfo
  {pages} {053501} (\bibinfo {year} {2016})}\BibitemShut {NoStop}%
\bibitem [{\citenamefont {Defienne}\ \emph
  {et~al.}(2018{\natexlab{a}})\citenamefont {Defienne}, \citenamefont
  {Reichert},\ and\ \citenamefont {Fleischer}}]{PhysRevLett.120.203604}%
  \BibitemOpen
  \bibfield  {author} {\bibinfo {author} {\bibfnamefont {H.}~\bibnamefont
  {Defienne}}, \bibinfo {author} {\bibfnamefont {M.}~\bibnamefont {Reichert}},\
  and\ \bibinfo {author} {\bibfnamefont {J.~W.}\ \bibnamefont {Fleischer}},\
  }\bibfield  {title} {\bibinfo {title} {{General Model of Photon-Pair
  Detection with an Image Sensor}},\ }\href
  {https://doi.org/10.1103/PhysRevLett.120.203604} {\bibfield  {journal}
  {\bibinfo  {journal} {Phys. Rev. Lett.}\ }\textbf {\bibinfo {volume} {120}},\
  \bibinfo {pages} {203604} (\bibinfo {year} {2018}{\natexlab{a}})}\BibitemShut
  {NoStop}%
\bibitem [{\citenamefont {Reichert}\ \emph {et~al.}(2018)\citenamefont
  {Reichert}, \citenamefont {Defienne},\ and\ \citenamefont
  {Fleischer}}]{PhysRevA.98.013841}%
  \BibitemOpen
  \bibfield  {author} {\bibinfo {author} {\bibfnamefont {M.}~\bibnamefont
  {Reichert}}, \bibinfo {author} {\bibfnamefont {H.}~\bibnamefont {Defienne}},\
  and\ \bibinfo {author} {\bibfnamefont {J.~W.}\ \bibnamefont {Fleischer}},\
  }\bibfield  {title} {\bibinfo {title} {Optimizing the signal-to-noise ratio
  of biphoton distribution measurements},\ }\href
  {https://doi.org/10.1103/PhysRevA.98.013841} {\bibfield  {journal} {\bibinfo
  {journal} {Phys. Rev. A}\ }\textbf {\bibinfo {volume} {98}},\ \bibinfo
  {pages} {013841} (\bibinfo {year} {2018})}\BibitemShut {NoStop}%
\bibitem [{\citenamefont {Ndagano}\ \emph {et~al.}(2022)\citenamefont
  {Ndagano}, \citenamefont {Defienne}, \citenamefont {Branford}, \citenamefont
  {Shah}, \citenamefont {Lyons}, \citenamefont {Westerberg}, \citenamefont
  {Gauger},\ and\ \citenamefont {Faccio}}]{Ndagano2022}%
  \BibitemOpen
  \bibfield  {author} {\bibinfo {author} {\bibfnamefont {B.}~\bibnamefont
  {Ndagano}}, \bibinfo {author} {\bibfnamefont {H.}~\bibnamefont {Defienne}},
  \bibinfo {author} {\bibfnamefont {D.}~\bibnamefont {Branford}}, \bibinfo
  {author} {\bibfnamefont {Y.~D.}\ \bibnamefont {Shah}}, \bibinfo {author}
  {\bibfnamefont {A.}~\bibnamefont {Lyons}}, \bibinfo {author} {\bibfnamefont
  {N.}~\bibnamefont {Westerberg}}, \bibinfo {author} {\bibfnamefont {E.~M.}\
  \bibnamefont {Gauger}},\ and\ \bibinfo {author} {\bibfnamefont
  {D.}~\bibnamefont {Faccio}},\ }\bibfield  {title} {\bibinfo {title} {{Quantum
  microscopy based on Hong--Ou--Mandel interference}},\ }\href
  {https://doi.org/10.1038/s41566-022-00980-6} {\bibfield  {journal} {\bibinfo
  {journal} {Nat. Photonics}\ }\textbf {\bibinfo {volume} {16}},\ \bibinfo
  {pages} {384} (\bibinfo {year} {2022})}\BibitemShut {NoStop}%
\bibitem [{\citenamefont {Chan}\ \emph {et~al.}(2007)\citenamefont {Chan},
  \citenamefont {Torres},\ and\ \citenamefont {Eberly}}]{PhysRevA.75.050101}%
  \BibitemOpen
  \bibfield  {author} {\bibinfo {author} {\bibfnamefont {K.~W.}\ \bibnamefont
  {Chan}}, \bibinfo {author} {\bibfnamefont {J.~P.}\ \bibnamefont {Torres}},\
  and\ \bibinfo {author} {\bibfnamefont {J.~H.}\ \bibnamefont {Eberly}},\
  }\bibfield  {title} {\bibinfo {title} {{Transverse entanglement migration in
  Hilbert space}},\ }\href {https://doi.org/10.1103/PhysRevA.75.050101}
  {\bibfield  {journal} {\bibinfo  {journal} {Phys. Rev. A}\ }\textbf {\bibinfo
  {volume} {75}},\ \bibinfo {pages} {050101(R)} (\bibinfo {year}
  {2007})}\BibitemShut {NoStop}%
\bibitem [{\citenamefont {Bhattacharjee}\ \emph {et~al.}(2022)\citenamefont
  {Bhattacharjee}, \citenamefont {Joshi}, \citenamefont {Karan}, \citenamefont
  {Leach},\ and\ \citenamefont {Jha}}]{Bhattacharjee2022}%
  \BibitemOpen
  \bibfield  {author} {\bibinfo {author} {\bibfnamefont {A.}~\bibnamefont
  {Bhattacharjee}}, \bibinfo {author} {\bibfnamefont {M.~K.}\ \bibnamefont
  {Joshi}}, \bibinfo {author} {\bibfnamefont {S.}~\bibnamefont {Karan}},
  \bibinfo {author} {\bibfnamefont {J.}~\bibnamefont {Leach}},\ and\ \bibinfo
  {author} {\bibfnamefont {A.~K.}\ \bibnamefont {Jha}},\ }\bibfield  {title}
  {\bibinfo {title} {{Propagation-induced revival of entanglement in the
  angle-OAM bases}},\ }\href {https://doi.org/10.1126/sciadv.abn7876}
  {\bibfield  {journal} {\bibinfo  {journal} {Sci. Adv.}\ }\textbf {\bibinfo
  {volume} {8}},\ \bibinfo {pages} {eabn7876} (\bibinfo {year}
  {2022})}\BibitemShut {NoStop}%
\bibitem [{\citenamefont {Law}\ and\ \citenamefont
  {Eberly}(2004)}]{PhysRevLett.92.127903}%
  \BibitemOpen
  \bibfield  {author} {\bibinfo {author} {\bibfnamefont {C.~K.}\ \bibnamefont
  {Law}}\ and\ \bibinfo {author} {\bibfnamefont {J.~H.}\ \bibnamefont
  {Eberly}},\ }\bibfield  {title} {\bibinfo {title} {{Analysis and
  Interpretation of High Transverse Entanglement in Optical Parametric Down
  Conversion}},\ }\href {https://doi.org/10.1103/PhysRevLett.92.127903}
  {\bibfield  {journal} {\bibinfo  {journal} {Phys. Rev. Lett.}\ }\textbf
  {\bibinfo {volume} {92}},\ \bibinfo {pages} {127903} (\bibinfo {year}
  {2004})}\BibitemShut {NoStop}%
\bibitem [{\citenamefont {Reichert}\ \emph {et~al.}(2017)\citenamefont
  {Reichert}, \citenamefont {Sun},\ and\ \citenamefont
  {Fleischer}}]{PhysRevA.95.063836}%
  \BibitemOpen
  \bibfield  {author} {\bibinfo {author} {\bibfnamefont {M.}~\bibnamefont
  {Reichert}}, \bibinfo {author} {\bibfnamefont {X.}~\bibnamefont {Sun}},\ and\
  \bibinfo {author} {\bibfnamefont {J.~W.}\ \bibnamefont {Fleischer}},\
  }\bibfield  {title} {\bibinfo {title} {Quality of spatial entanglement
  propagation},\ }\href {https://doi.org/10.1103/PhysRevA.95.063836} {\bibfield
   {journal} {\bibinfo  {journal} {Phys. Rev. A}\ }\textbf {\bibinfo {volume}
  {95}},\ \bibinfo {pages} {063836} (\bibinfo {year} {2017})}\BibitemShut
  {NoStop}%
\bibitem [{\citenamefont {Saleh}\ \emph {et~al.}(2005)\citenamefont {Saleh},
  \citenamefont {Teich},\ and\ \citenamefont
  {Sergienko}}]{PhysRevLett.94.223601}%
  \BibitemOpen
  \bibfield  {author} {\bibinfo {author} {\bibfnamefont {B.~E.~A.}\
  \bibnamefont {Saleh}}, \bibinfo {author} {\bibfnamefont {M.~C.}\ \bibnamefont
  {Teich}},\ and\ \bibinfo {author} {\bibfnamefont {A.~V.}\ \bibnamefont
  {Sergienko}},\ }\bibfield  {title} {\bibinfo {title} {{Wolf Equations for
  Two-Photon Light}},\ }\href {https://doi.org/10.1103/PhysRevLett.94.223601}
  {\bibfield  {journal} {\bibinfo  {journal} {Phys. Rev. Lett.}\ }\textbf
  {\bibinfo {volume} {94}},\ \bibinfo {pages} {223601} (\bibinfo {year}
  {2005})}\BibitemShut {NoStop}%
\bibitem [{\citenamefont {Defienne}\ \emph
  {et~al.}(2018{\natexlab{b}})\citenamefont {Defienne}, \citenamefont
  {Reichert},\ and\ \citenamefont {Fleischer}}]{PhysRevLett.121.233601}%
  \BibitemOpen
  \bibfield  {author} {\bibinfo {author} {\bibfnamefont {H.}~\bibnamefont
  {Defienne}}, \bibinfo {author} {\bibfnamefont {M.}~\bibnamefont {Reichert}},\
  and\ \bibinfo {author} {\bibfnamefont {J.~W.}\ \bibnamefont {Fleischer}},\
  }\bibfield  {title} {\bibinfo {title} {{Adaptive Quantum Optics with
  Spatially Entangled Photon Pairs}},\ }\href
  {https://doi.org/10.1103/PhysRevLett.121.233601} {\bibfield  {journal}
  {\bibinfo  {journal} {Phys. Rev. Lett.}\ }\textbf {\bibinfo {volume} {121}},\
  \bibinfo {pages} {233601} (\bibinfo {year} {2018}{\natexlab{b}})}\BibitemShut
  {NoStop}%
\bibitem [{\citenamefont {Cameron}\ \emph {et~al.}(2024)\citenamefont
  {Cameron}, \citenamefont {Courme}, \citenamefont {Vernière}, \citenamefont
  {Pandya}, \citenamefont {Faccio},\ and\ \citenamefont
  {Defienne}}]{science.adk7825}%
  \BibitemOpen
  \bibfield  {author} {\bibinfo {author} {\bibfnamefont {P.}~\bibnamefont
  {Cameron}}, \bibinfo {author} {\bibfnamefont {B.}~\bibnamefont {Courme}},
  \bibinfo {author} {\bibfnamefont {C.}~\bibnamefont {Vernière}}, \bibinfo
  {author} {\bibfnamefont {R.}~\bibnamefont {Pandya}}, \bibinfo {author}
  {\bibfnamefont {D.}~\bibnamefont {Faccio}},\ and\ \bibinfo {author}
  {\bibfnamefont {H.}~\bibnamefont {Defienne}},\ }\bibfield  {title} {\bibinfo
  {title} {Adaptive optical imaging with entangled photons},\ }\href
  {https://doi.org/10.1126/science.adk7825} {\bibfield  {journal} {\bibinfo
  {journal} {Science}\ }\textbf {\bibinfo {volume} {383}},\ \bibinfo {pages}
  {1142} (\bibinfo {year} {2024})}\BibitemShut {NoStop}%
\bibitem [{\citenamefont {Hradil}\ \emph {et~al.}(2010)\citenamefont {Hradil},
  \citenamefont {\ifmmode \check{R}\else \v{R}\fi{}eh\'a\ifmmode~\check{c}\else
  \v{c}\fi{}ek},\ and\ \citenamefont
  {S\'anchez-Soto}}]{PhysRevLett.105.010401}%
  \BibitemOpen
  \bibfield  {author} {\bibinfo {author} {\bibfnamefont {Z.}~\bibnamefont
  {Hradil}}, \bibinfo {author} {\bibfnamefont {J.}~\bibnamefont {\ifmmode
  \check{R}\else \v{R}\fi{}eh\'a\ifmmode~\check{c}\else \v{c}\fi{}ek}},\ and\
  \bibinfo {author} {\bibfnamefont {L.~L.}\ \bibnamefont {S\'anchez-Soto}},\
  }\bibfield  {title} {\bibinfo {title} {{Quantum Reconstruction of the Mutual
  Coherence Function}},\ }\href
  {https://doi.org/10.1103/PhysRevLett.105.010401} {\bibfield  {journal}
  {\bibinfo  {journal} {Phys. Rev. Lett.}\ }\textbf {\bibinfo {volume} {105}},\
  \bibinfo {pages} {010401} (\bibinfo {year} {2010})}\BibitemShut {NoStop}%
\bibitem [{\citenamefont {Stoklasa}\ \emph {et~al.}(2014)\citenamefont
  {Stoklasa}, \citenamefont {Motka}, \citenamefont {Rehacek}, \citenamefont
  {Hradil},\ and\ \citenamefont {S{\'a}nchez-Soto}}]{Stoklasa2014}%
  \BibitemOpen
  \bibfield  {author} {\bibinfo {author} {\bibfnamefont {B.}~\bibnamefont
  {Stoklasa}}, \bibinfo {author} {\bibfnamefont {L.}~\bibnamefont {Motka}},
  \bibinfo {author} {\bibfnamefont {J.}~\bibnamefont {Rehacek}}, \bibinfo
  {author} {\bibfnamefont {Z.}~\bibnamefont {Hradil}},\ and\ \bibinfo {author}
  {\bibfnamefont {L.~L.}\ \bibnamefont {S{\'a}nchez-Soto}},\ }\bibfield
  {title} {\bibinfo {title} {Wavefront sensing reveals optical coherence},\
  }\href {https://doi.org/10.1038/ncomms4275} {\bibfield  {journal} {\bibinfo
  {journal} {Nat. Commun.}\ }\textbf {\bibinfo {volume} {5}},\ \bibinfo {pages}
  {3275} (\bibinfo {year} {2014})}\BibitemShut {NoStop}%
\bibitem [{\citenamefont {Zhou}\ \emph {et~al.}(2021)\citenamefont {Zhou},
  \citenamefont {Zhao}, \citenamefont {Hay}, \citenamefont {McGonagle},
  \citenamefont {Boyd},\ and\ \citenamefont {Shi}}]{PhysRevLett.127.040402}%
  \BibitemOpen
  \bibfield  {author} {\bibinfo {author} {\bibfnamefont {Y.}~\bibnamefont
  {Zhou}}, \bibinfo {author} {\bibfnamefont {J.}~\bibnamefont {Zhao}}, \bibinfo
  {author} {\bibfnamefont {D.}~\bibnamefont {Hay}}, \bibinfo {author}
  {\bibfnamefont {K.}~\bibnamefont {McGonagle}}, \bibinfo {author}
  {\bibfnamefont {R.~W.}\ \bibnamefont {Boyd}},\ and\ \bibinfo {author}
  {\bibfnamefont {Z.}~\bibnamefont {Shi}},\ }\bibfield  {title} {\bibinfo
  {title} {{Direct Tomography of High-Dimensional Density Matrices for General
  Quantum States of Photons}},\ }\href
  {https://doi.org/10.1103/PhysRevLett.127.040402} {\bibfield  {journal}
  {\bibinfo  {journal} {Phys. Rev. Lett.}\ }\textbf {\bibinfo {volume} {127}},\
  \bibinfo {pages} {040402} (\bibinfo {year} {2021})}\BibitemShut {NoStop}%
\bibitem [{gra()}]{gradientnote}%
  \BibitemOpen
  \href@noop {} {\bibinfo {title} {{The symbol $\nabla_1$ means taking the
  gradient with respect to the first vector variable. For example,
  $\nabla_1f(\boldsymbol{\rho},\boldsymbol{\rho})=\nabla_{\boldsymbol{\rho}'}f(\boldsymbol{\rho}',\boldsymbol{\rho})|_{\boldsymbol{\rho}'=\boldsymbol{\rho}}$.}}}\BibitemShut
  {Stop}%
\bibitem [{\citenamefont {Goodman}(2017)}]{FourierOptics}%
  \BibitemOpen
  \bibfield  {author} {\bibinfo {author} {\bibfnamefont {J.~W.}\ \bibnamefont
  {Goodman}},\ }\href@noop {} {\emph {\bibinfo {title} {Introduction to Fourier
  Optics}}},\ \bibinfo {edition} {4th}\ ed.\ (\bibinfo  {publisher} {W. H.
  Freeman},\ \bibinfo {year} {2017})\BibitemShut {NoStop}%
\bibitem [{\citenamefont {Eimerl}\ \emph {et~al.}(1987)\citenamefont {Eimerl},
  \citenamefont {Davis}, \citenamefont {Velsko}, \citenamefont {Graham},\ and\
  \citenamefont {Zalkin}}]{10.1063/1.339536}%
  \BibitemOpen
  \bibfield  {author} {\bibinfo {author} {\bibfnamefont {D.}~\bibnamefont
  {Eimerl}}, \bibinfo {author} {\bibfnamefont {L.}~\bibnamefont {Davis}},
  \bibinfo {author} {\bibfnamefont {S.}~\bibnamefont {Velsko}}, \bibinfo
  {author} {\bibfnamefont {E.~K.}\ \bibnamefont {Graham}},\ and\ \bibinfo
  {author} {\bibfnamefont {A.}~\bibnamefont {Zalkin}},\ }\bibfield  {title}
  {\bibinfo {title} {{Optical, mechanical, and thermal properties of barium
  borate}},\ }\href {https://doi.org/10.1063/1.339536} {\bibfield  {journal}
  {\bibinfo  {journal} {J. Appl. Phys.}\ }\textbf {\bibinfo {volume} {62}},\
  \bibinfo {pages} {1968} (\bibinfo {year} {1987})}\BibitemShut {NoStop}%
\bibitem [{\citenamefont {Courme}\ \emph {et~al.}(2023)\citenamefont {Courme},
  \citenamefont {Verni\`{e}re}, \citenamefont {Svihra}, \citenamefont {Gigan},
  \citenamefont {Nomerotski},\ and\ \citenamefont {Defienne}}]{Courme:23}%
  \BibitemOpen
  \bibfield  {author} {\bibinfo {author} {\bibfnamefont {B.}~\bibnamefont
  {Courme}}, \bibinfo {author} {\bibfnamefont {C.}~\bibnamefont
  {Verni\`{e}re}}, \bibinfo {author} {\bibfnamefont {P.}~\bibnamefont
  {Svihra}}, \bibinfo {author} {\bibfnamefont {S.}~\bibnamefont {Gigan}},
  \bibinfo {author} {\bibfnamefont {A.}~\bibnamefont {Nomerotski}},\ and\
  \bibinfo {author} {\bibfnamefont {H.}~\bibnamefont {Defienne}},\ }\bibfield
  {title} {\bibinfo {title} {Quantifying high-dimensional spatial entanglement
  with a single-photon-sensitive time-stamping camera},\ }\href
  {https://doi.org/10.1364/OL.487182} {\bibfield  {journal} {\bibinfo
  {journal} {Opt. Lett.}\ }\textbf {\bibinfo {volume} {48}},\ \bibinfo {pages}
  {3439} (\bibinfo {year} {2023})}\BibitemShut {NoStop}%
\bibitem [{\citenamefont {Einstein}\ \emph {et~al.}(1935)\citenamefont
  {Einstein}, \citenamefont {Podolsky},\ and\ \citenamefont
  {Rosen}}]{EPRpaper}%
  \BibitemOpen
  \bibfield  {author} {\bibinfo {author} {\bibfnamefont {A.}~\bibnamefont
  {Einstein}}, \bibinfo {author} {\bibfnamefont {B.}~\bibnamefont {Podolsky}},\
  and\ \bibinfo {author} {\bibfnamefont {N.}~\bibnamefont {Rosen}},\ }\bibfield
   {title} {\bibinfo {title} {{Can Quantum-Mechanical Description of Physical
  Reality Be Considered Complete?}},\ }\href
  {https://doi.org/10.1103/PhysRev.47.777} {\bibfield  {journal} {\bibinfo
  {journal} {Phys. Rev.}\ }\textbf {\bibinfo {volume} {47}},\ \bibinfo {pages}
  {777} (\bibinfo {year} {1935})}\BibitemShut {NoStop}%
\bibitem [{\citenamefont {Yi}\ \emph {et~al.}(2021)\citenamefont {Yi},
  \citenamefont {Xiang}, \citenamefont {Zhou}, \citenamefont {Wu},
  \citenamefont {Yang},\ and\ \citenamefont {Yu}}]{yi2021angle}%
  \BibitemOpen
  \bibfield  {author} {\bibinfo {author} {\bibfnamefont {S.}~\bibnamefont
  {Yi}}, \bibinfo {author} {\bibfnamefont {J.}~\bibnamefont {Xiang}}, \bibinfo
  {author} {\bibfnamefont {M.}~\bibnamefont {Zhou}}, \bibinfo {author}
  {\bibfnamefont {Z.}~\bibnamefont {Wu}}, \bibinfo {author} {\bibfnamefont
  {L.}~\bibnamefont {Yang}},\ and\ \bibinfo {author} {\bibfnamefont
  {Z.}~\bibnamefont {Yu}},\ }\bibfield  {title} {\bibinfo {title} {Angle-based
  wavefront sensing enabled by the near fields of flat optics},\ }\href
  {https://doi.org/10.1038/s41467-021-26169-z} {\bibfield  {journal} {\bibinfo
  {journal} {Nat. Commun.}\ }\textbf {\bibinfo {volume} {12}},\ \bibinfo
  {pages} {6002} (\bibinfo {year} {2021})}\BibitemShut {NoStop}%
\bibitem [{\citenamefont {Nirala}\ \emph {et~al.}(2023)\citenamefont {Nirala},
  \citenamefont {Pradyumna}, \citenamefont {Kumar},\ and\ \citenamefont
  {Marino}}]{doi:10.1126/sciadv.adf9161}%
  \BibitemOpen
  \bibfield  {author} {\bibinfo {author} {\bibfnamefont {G.}~\bibnamefont
  {Nirala}}, \bibinfo {author} {\bibfnamefont {S.~T.}\ \bibnamefont
  {Pradyumna}}, \bibinfo {author} {\bibfnamefont {A.}~\bibnamefont {Kumar}},\
  and\ \bibinfo {author} {\bibfnamefont {A.~M.}\ \bibnamefont {Marino}},\
  }\bibfield  {title} {\bibinfo {title} {Information encoding in the spatial
  correlations of entangled twin beams},\ }\href
  {https://doi.org/10.1126/sciadv.adf9161} {\bibfield  {journal} {\bibinfo
  {journal} {Sci. Adv.}\ }\textbf {\bibinfo {volume} {9}},\ \bibinfo {pages}
  {eadf9161} (\bibinfo {year} {2023})}\BibitemShut {NoStop}%
\bibitem [{\citenamefont {Liu}\ \emph {et~al.}(2018)\citenamefont {Liu},
  \citenamefont {Lyyra}, \citenamefont {Sun}, \citenamefont {Liu},
  \citenamefont {Li}, \citenamefont {Guo}, \citenamefont {Maniscalco},\ and\
  \citenamefont {Piilo}}]{Liu2018}%
  \BibitemOpen
  \bibfield  {author} {\bibinfo {author} {\bibfnamefont {Z.-D.}\ \bibnamefont
  {Liu}}, \bibinfo {author} {\bibfnamefont {H.}~\bibnamefont {Lyyra}}, \bibinfo
  {author} {\bibfnamefont {Y.-N.}\ \bibnamefont {Sun}}, \bibinfo {author}
  {\bibfnamefont {B.-H.}\ \bibnamefont {Liu}}, \bibinfo {author} {\bibfnamefont
  {C.-F.}\ \bibnamefont {Li}}, \bibinfo {author} {\bibfnamefont {G.-C.}\
  \bibnamefont {Guo}}, \bibinfo {author} {\bibfnamefont {S.}~\bibnamefont
  {Maniscalco}},\ and\ \bibinfo {author} {\bibfnamefont {J.}~\bibnamefont
  {Piilo}},\ }\bibfield  {title} {\bibinfo {title} {Experimental implementation
  of fully controlled dephasing dynamics and synthetic spectral densities},\
  }\href {https://doi.org/10.1038/s41467-018-05817-x} {\bibfield  {journal}
  {\bibinfo  {journal} {Nat. Commun.}\ }\textbf {\bibinfo {volume} {9}},\
  \bibinfo {pages} {3453} (\bibinfo {year} {2018})}\BibitemShut {NoStop}%
\bibitem [{\citenamefont {Liu}\ \emph {et~al.}(2020)\citenamefont {Liu},
  \citenamefont {Sun}, \citenamefont {Liu}, \citenamefont {Li}, \citenamefont
  {Guo}, \citenamefont {Hamedani~Raja}, \citenamefont {Lyyra},\ and\
  \citenamefont {Piilo}}]{PhysRevA.102.062208}%
  \BibitemOpen
  \bibfield  {author} {\bibinfo {author} {\bibfnamefont {Z.-D.}\ \bibnamefont
  {Liu}}, \bibinfo {author} {\bibfnamefont {Y.-N.}\ \bibnamefont {Sun}},
  \bibinfo {author} {\bibfnamefont {B.-H.}\ \bibnamefont {Liu}}, \bibinfo
  {author} {\bibfnamefont {C.-F.}\ \bibnamefont {Li}}, \bibinfo {author}
  {\bibfnamefont {G.-C.}\ \bibnamefont {Guo}}, \bibinfo {author} {\bibfnamefont
  {S.}~\bibnamefont {Hamedani~Raja}}, \bibinfo {author} {\bibfnamefont
  {H.}~\bibnamefont {Lyyra}},\ and\ \bibinfo {author} {\bibfnamefont
  {J.}~\bibnamefont {Piilo}},\ }\bibfield  {title} {\bibinfo {title}
  {{Experimental realization of high-fidelity teleportation via a non-Markovian
  open quantum system}},\ }\href {https://doi.org/10.1103/PhysRevA.102.062208}
  {\bibfield  {journal} {\bibinfo  {journal} {Phys. Rev. A}\ }\textbf {\bibinfo
  {volume} {102}},\ \bibinfo {pages} {062208} (\bibinfo {year}
  {2020})}\BibitemShut {NoStop}%
\bibitem [{\citenamefont {Tam}\ \emph {et~al.}(2022)\citenamefont {Tam},
  \citenamefont {Claassen},\ and\ \citenamefont {Kane}}]{PhysRevX.12.031022}%
  \BibitemOpen
  \bibfield  {author} {\bibinfo {author} {\bibfnamefont {P.~M.}\ \bibnamefont
  {Tam}}, \bibinfo {author} {\bibfnamefont {M.}~\bibnamefont {Claassen}},\ and\
  \bibinfo {author} {\bibfnamefont {C.~L.}\ \bibnamefont {Kane}},\ }\bibfield
  {title} {\bibinfo {title} {{Topological Multipartite Entanglement in a Fermi
  Liquid}},\ }\href {https://doi.org/10.1103/PhysRevX.12.031022} {\bibfield
  {journal} {\bibinfo  {journal} {Phys. Rev. X}\ }\textbf {\bibinfo {volume}
  {12}},\ \bibinfo {pages} {031022} (\bibinfo {year} {2022})}\BibitemShut
  {NoStop}%
\end{thebibliography}
\end{document}